\documentclass[]{aa} 
\pdfoutput=1
\makeatletter
\renewcommand*\aa@pageof{, page \thepage{} of \pageref*{LastPage}}
\makeatother
\usepackage{graphicx,xspace}
\usepackage{siunitx}[=v2]
\newcommand{\ngc}{NGC\,7635\xspace}
\newcommand{\bdforty}{BD+43$^\circ$\,3654\xspace}
\newcommand{\bdsixty}{BD+60$^\circ$\,2522\xspace}

\usepackage[acronym]{glossaries}
\setacronymstyle{long-short}
\makeglossaries
\newacronym{ism}{ISM}{interstellar medium}
\newacronym{ir}{IR}{infrared}
\usepackage[title]{appendix}
\usepackage{grffile,txfonts,twoopt}
\usepackage[colorlinks=true, allcolors=blue,breaklinks=true]{hyperref}
\bibpunct{(}{)}{;}{a}{}{,}             %% natbib format for A&A and ApJ
\makeatletter
  \newcommandtwoopt{\citeads}[3][][]{\href{http://adsabs.harvard.edu/abs/#3}%
    {\def\hyper@linkstart##1##2{}%
     \let\hyper@linkend\@empty\citealp[#1][#2]{#3}}}
     
  \newcommandtwoopt{\citepads}[3][][]{\href{http://adsabs.harvard.edu/abs/#3}%
    {\def\hyper@linkstart##1##2{}%
     \let\hyper@linkend\@empty\citep[#1][#2]{#3}}}
     
 \newcommandtwoopt{\citeauthorads}[3][][]{\href{http://adsabs.harvard.edu/abs/#3}%
    {\def\hyper@linkstart##1##2{}%
     \let\hyper@linkend\@empty\citeauthor[#1][#2]{#3}}}
     
  \newcommandtwoopt{\citeadsalias}[3][][]{\href{http://adsabs.harvard.edu/abs/#3}%
    {\def\hyper@linkstart##1##2{}%
     \let\hyper@linkend\@empty\citetalias[#1][#2]{#3}}}     

  \newcommandtwoopt{\citetads}[3][][]{\href{http://adsabs.harvard.edu/abs/#3}%
    {\def\hyper@linkstart##1##2{}%
     \let\hyper@linkend\@empty\citet[#1][#2]{#3}}}
  \newcommandtwoopt{\citeyearads}[3][][]%
    {\href{http://adsabs.harvard.edu/abs/#3}
    {\def\hyper@linkstart##1##2{}%
     \let\hyper@linkend\@empty\citeyear[#1][#2]{#3}}}
\makeatother
\defcitealias{2010A&A...517L..10B}{B10}
\defcitealias{2021MNRAS.503.2514B}{B21}
\usepackage{booktabs,caption,color,soul}
\usepackage[flushleft]{threeparttable}
\usepackage[normalem]{ulem}

\makeatletter
\newcommand\footnoteref[1]{\protected@xdef\@thefnmark{\ref{#1}}\@footnotemark}
\makeatother
\begin{document} 
%%%%%%%%%%%%%%%%%%%%%%%%%%%%%%%%%%%%%%%
% TITLE
%%%%%%%%%%%%%%%%%%%%%%%%%%%%%%%%%%%%%%%
\title{And then they were two: detection of non-thermal radio emission from the bow shocks of two runaway stars}

\author{M.~Moutzouri \inst{\ref{inst1} \thanks{\email{moutzouri@cp.dias.ie}}
          ,\ref{inst2}}
      \and
      J.~Mackey\inst{\ref{inst1}}
      \and
      C.~Carrasco-Gonz\'alez\inst{\ref{inst3}}
      \and
      Y.~Gong\inst{\ref{inst4}}
      \and
      R.~Brose\inst{\ref{inst1}}
      \and
      D.~Zargaryan\inst{\ref{inst1}}
      \and
      J.~A.~Toal\'a\inst{\ref{inst3}}
      \and
      K.~M.~Menten\inst{\ref{inst4}}
      \and
      V.~V.~Gvaramadze\inst{\ref{inst5},\ref{inst6},\ref{inst7}}
      \and
      M.~R.~Rugel\inst{\ref{inst4}}
    }

\institute{Dublin Institute for Advanced Studies (DIAS), 31 Fitzwilliam Place, Dublin 2, Ireland \label{inst1}
   \and
   School of Physics, University College Dublin, Belfield, Dublin 4, Ireland \label{inst2}
   \and
   Instituto de Radioastronom\'ia y Astrof\'isica (IRyA), Universidad Nacional Aut\'onoma de M\'exico, Morelia, 58089, Mexico \label{inst3}
   \and
   Max-Planck-Institut f\"ur Radioastronomie (MPIfR), Auf dem H\"ugel 69, D-
53121 Bonn, Germany \label{inst4}
   \and
   Sternberg Astronomical Institute, Lomonosov Moscow State University, Universitetskij Pr. 13, Moscow 119992, Russia \label{inst5}
   \and
   Space Research Institute, Russian Academy of Sciences, Profsoyuznaya 84/32, Moscow 117997, Russia \label{inst6}
   \and
   E.~Kharadze Georgian National Astrophysical Observatory, Abastumani 0301, Georgia \label{inst7}
}

%%%%%%%%%%%%%%%%%%%%%%%%%%%%%%%%%%%%%%%
% ABSTRACT
%%%%%%%%%%%%%%%%%%%%%%%%%%%%%%%%%%%%%%%
\abstract
% context
{In recent years, winds from massive stars have been considered promising sites for investigating relativistic particle acceleration. 
In particular, the resulting bow-shaped shock from the interaction of the supersonic winds of runaway stars with interstellar matter have been intensively observed at many different wavelengths from radio to $\gamma$-rays.}
% aims 
{In this study we investigate the O4If star, \bdforty, the bow shock of which is, so far, the only one proven to radiate both thermal and non-thermal emission at radio frequencies. 
In addition, we consider \ngc, the Bubble Nebula, as a bow shock candidate and examine its apex for indications of thermal and non-thermal radio emission.}
% methods 
{We observed both bow shocks in radio frequencies with the Very Large Array in C and X bands (\SIrange{4}{8}{\GHz} and \SIrange{8}{12}{GHz}), and with the Effelsberg telescope at \SIrange{4}{8}{\GHz}. 
We analysed single-dish and interferometric results individually, in addition to their combined emission, obtained spectral index maps for each source and calculated their Spectral Energy Distributions.}
% results 
{We find that both sources emit non-thermal emission in the radio regime, with the clearest evidence for \ngc, whose radio emission has a strongly negative spectral index along the northern rim of the bubble.
We present the first high-resolution maps of radio emission from \ngc, finding that the morphology follows closely the optical nebular emission.
Our results are less conclusive for the bow shock of \bdforty, as its emission becomes weaker and faint at higher frequencies in VLA data.
Effelsberg data shows a much larger emitting region (albeit thermal emission) than is detected with the VLA for this source.}
% conclusions 
{Our results extend the previous radio results from the \bdforty bow shock to higher frequencies, and with our \ngc results we double the number of bow shocks around O stars with detected non-thermal emission from one to two.
Modelling of the multi-wavelength data for both sources shows that accelerated electrons at the wind termination shock are a plausible source for the non-thermal radio emission, but energetics arguments suggest that any non-thermal X-ray and $\gamma$-ray emission could be significantly below existing upper limits.
Enhanced synchrotron emission from compressed Galactic cosmic rays in the radiative bow shock could also explain the radio emission from the \bdforty bow shock but not \ngc.
Non-detection of point-like radio emission from \bdforty puts an upper limit on the mass-loss rate of the star that is lower than values quoted in the literature.} 

\titlerunning{Non-thermal radio emission from bow shocks}
\authorrunning{Moutzouri et al.} 
 
% \object{BD+43 3654}
% \object{NGC 7635}
% \object{BD+60 2522}
\maketitle
%%%%%%%%%%%%%%%%%%%%%%%%%%%%%%%%%%%%%%%
% 01 - INTRODUCTION
%%%%%%%%%%%%%%%%%%%%%%%%%%%%%%%%%%%%%%%
\section{Introduction}\label{sec:intro}
OB-runaway stars are an unusual group of stars  \citepads[about only  20\% of O- and B-types,][]{2011Sci...334.1380F} that are ejected from their parent clusters and travel through the \gls{ism} with stellar velocities, $v_\star$, of more than \SI{30}{km.s^{-1}}, relative to the material they move through.
Their strong winds, which have terminal velocities, $v_\infty$, varying from 1000 to \SI{3000}{km.s^{-1}}, interact with the \gls{ism} hydrodynamically \citepads{1977ApJ...218..377W}.
In particular, when $v_\star>c_\mathrm{s}$ ($c_\mathrm{s}$ is the sound speed of the \gls{ism} the star is passing through) the interaction of the stellar wind with the interstellar matter sweeps up and compresses it, creating an impressive, arc-shaped nebula, often referred to as a bow shock (\citeads{1979ApJ...230..782G}, \citeads{1988ApJ...329L..93V}). 

Bow shocks emit predominantly at mid-\gls{ir} wavelengths through thermal emission from dust that is radiatively heated by the wind-driving star (\citeads{1988ApJ...329L..93V}, \citeads{2014MNRAS.444.2754M}, \citeads{2019MNRAS.489.2142H}).
\gls{ir} observations can, however, be difficult to interpret because they trace the dust and not the gas directly, and so the uncertainties in obtaining the gas density distribution are not small.
The mid-\gls{ir} brightness is very sensitive to the assumed grain size-distribution and composition (\citeads{2013ARep...57..573P}, \citeads{2016A&A...586A.114M}, \citeads{2019MNRAS.486.4947K}).
The emission scales with the dust density, and there are regions of parameter space where it is expected that dust and gas are not strongly coupled and the dust is a poor tracer of gas (\citeads{2015MNRAS.449..440A}, \citeads{2018MNRAS.473.1576K}, \citeads{2019MNRAS.486.4423H}).

Emission from nebular optical lines, such as H$\alpha$ or [O\,\textsc{iii}], is fainter \citepads{1979ApJ...230..782G} but a more reliable tracer of the gas conditions in the bow shock.
Unfortunately, in this case, the stars driving bow shock nebulae tend to be very bright themselves, which may cause problems when trying to observe faint nebular emission.
Furthermore, optical emission is significantly affected by interstellar extinction that is often poorly constrained and sometimes patchy across an extended nebula \citepads{2009A&A...496..177D}.

Radio bremsstrahlung, on the other hand, is potentially a better observational tracer of the gas density because it is unaffected by extinction and because massive stars themselves are relatively faint radio sources.
Radio can also be the best frequency range at which to detect diffuse non-thermal synchrotron radiation, which can be brighter than thermal bremsstrahlung at low frequencies \citepads[e.g.][]{2010A&A...517L..10B}.
This possibility to detect both thermal and non-thermal emission from bow shocks with radio observations, can provide valuable constraints on the physical conditions of the thermal and relativistic particles in a bow shock \citepads{2018A&A...617A..13D}, but one must first characterise the nature of the emission and separate TE from non-thermal (NTE)\footnote{In the following, TE and NTE refer to thermal and non-thermal \textit{radio} emission, respectively, unless noted otherwise.}.

NTE (synchrotron) arises from particles that are accelerated to relativistic energies in the termination shock of the stellar wind via Diffusive Shock Acceleration \citepads[DSA,][]{1983RPPh...46..973D} and potentially also by synchrotron radiation from Galactic Cosmic Rays (GCRs) compressed and deflected by the swept up interstellar magnetic field in the bow shock \citepads{2019A&A...622A..57C}.
Following early theoretical calculations by \citetads{1983SSRv...36..173C},
there has recently been renewed interest in the possibility that winds from massive stars contribute significantly to the very high-energy GCR population \citepads{2019NatAs...3..561A}.
Of all the massive star systems without compact-object companions, only $\eta$ Carinae (\citeads{2009ApJ...698L.142T}, \citeads{2020A&A...635A.167H}) and $\gamma^2$ Velorum (WR\,11, \citeads{2020A&A...635A.141M}, \citeads{2016MNRAS.457L..99P}), both colliding-wind binaries, have been detected in $\gamma$-rays.

So far, there has been no detection of NTE from bow shocks at very high energies \citepads[e.g.][]{2018A&A...612A..12H}. 
Similarly, only upper limits exist in X-rays (\citeads{2016ApJ...821...79T}, \citeads{2017ApJ...838L..19T}). 
Nevertheless, at radio frequencies, there is the unique case of the bow shock associated with the star \bdforty, from where previous studies (\citeads{2010A&A...517L..10B}, \citeads{2021MNRAS.503.2514B} from now on \citeadsalias{2010A&A...517L..10B}, \citeadsalias{2021MNRAS.503.2514B}) detect both TE and NTE at low GHz frequencies.
Following the detection of synchrotron radiation from the Wolf-Rayet nebula G2.4+1.4 around WR\,102 \citepads{2019ApJ...884L..49P}, and TE/NTE from the bow shock of the high-mass X-ray binary, Vela X-1 \citepads{2022MNRAS.510..515V}, it becomes clear that radio observations are a promising avenue to identify candidate bow shocks with strong NTE that can then be followed up at higher energies.
What is more, the well-known Bubble Nebula around the star \bdsixty was recently suggested to be a bow shock candidate by \citetads{2019A&A...625A...4G}. 
Given that this source is very bright and has similar properties to the bow shock of \bdforty, it was noted to be another good candidate for detecting both TE and NTE, although X-ray observations again only resulted in upper limits \citepads{2020MNRAS.495.3041T}.

Motivated by these results, as well as the extended capabilities of the National Science Foundation's (NSF) Karl G. Jansky Very Large Array  \citepads[VLA,][]{2011ApJ...739L...1P}, operated by the US National Radio Astronomy Observatory (NRAO), we investigate the bow shock of \bdforty and the nebula of \bdsixty with radio observations at \SIrange{4}{12}{\GHz}, applying several techniques to analyse their spectra.
We extend the spectral coverage of previous \bdforty observations (\citeadsalias{2010A&A...517L..10B}, \citeadsalias{2021MNRAS.503.2514B}) to higher frequencies, where we expect TE to become more prominent. 
We also present the first high-resolution radio maps and spectral analysis of the nebula \ngc around \bdsixty.

Given the difficulties in imaging extended diffuse sources with interferometers, it is important to obtain wide frequency coverage to more robustly characterise the thermal and non-thermal contributions to the observed emission.
It is also important to assess how much flux is lost due to lack of sensitivity to large scale uniform emission. 
For this reason we have supplemented our VLA observations with single-dish observations made with the Effelsberg 100 meter radio telescope at \SIrange{4}{8}{\GHz}.

In this paper, we present the results from our observations conducted with the VLA at the frequencies \SIrange{4}{12}{\GHz}, along with single-dish data obtained with the Effelsberg telescope at similar frequencies, and finally a combination of both.
In Section \ref{sec:tar} we present our targets in detail, while in Section \ref{sec:obs} we describe the observations; in Section \ref{sec:rea} the results of the interferometric observations are presented; the single-dish observations and the combined results are shown in Section \ref{subs:comb}.
We discuss our results in Section \ref{sec:disc} and, finally, our conclusions are presented in Section \ref{sec:concl}.

%%%%%%%%%%%%%%%%%%%%%%%%%%%%%%%%%%%%%%%
% 02 - TARGETS
%%%%%%%%%%%%%%%%%%%%%%%%%%%%%%%%%%%%%%%
\section{Targets}\label{sec:tar}
The bow shock around the massive star \bdforty and the nebula \ngc around the star \bdsixty are both wind-blown nebulae producing bow shocks and, as such, the properties of the central stars are important for determining the observational characteristics of the nebulae.
Some relevant properties of both stars taken from literature are shown in Table~\ref{tab:lit_info}.

%%%%%%%%%%%%%%%%%%%%%%%%%%%%%%%%%%%%%%%
%%%%%%%%%%%%%%%%%%%%%%%%%%%%%%%%%%%%%%%
\begin{table}
    \caption{Parameters of the massive stars driving the two bow shocks.
    } \label{tab:lit_info}
    \begin{threeparttable}
    \centering
    \vspace{-0.25cm}
    %%%%%%%%%%%%%%%%%%%%%
    \begin{tabular}{lrrr}
    %%%%%%%%%%%%%%%%%%%%%
    \toprule
    \midrule
    %%%%%%%%%%%%%%%%%%%%%
    & \bdforty 
    & \bdsixty 
    & Refs. \\
    \midrule
    %%%%%%%%%%%%%%%%%%%%%
    R.A.$_\textrm{J2000}$             
    & $20^{\rm h} 33^{\rm m} 36\rlap{.}^{\rm s} 8$ %20h 33m 36.8s  
    & $23^{\rm h} 20^{\rm m} 44\rlap{.}^{\rm s} 5$ 
    & (1) \\
    %%%%%%%%%%%%%%%%%%%%%
    Dec.$_\textrm{J2000}$
    & +\ang{43;59;07.4}
    & +\ang{61;11;40.5} 
    & (1) \\
    %%%%%%%%%%%%%%%%%%%%%
    spec. type 
    & O4If       
    & O6.5(n)fp  
    & (2),(3) \\
    %%%%%%%%%%%%%%%%%%%%%
    $d$ (kpc) 
    & $1.72\pm0.03$    
    & $3.0\pm0.2$    
    & (1) \\
    %%%%%%%%%%%%%%%%%%%%%
    $M_\star$ (M$_\odot$) 
    & $70\pm15$               
    & $39\pm10$                
    & (4),(5) \\
    %%%%%%%%%%%%%%%%%%%%%
    $v_\mathrm{tr}$ (\si{km.s^{-1}}) 
    & $43\pm1$                 
    & $28\pm1$                  
    & (1) \\
    %%%%%%%%%%%%%%%%%%%%%
    $v_\mathrm{r}$ (\si{km.s^{-1}}) 
    & [$-54,90$]                 
    & [$6,28$]
    & (6),(7) \\
    %%%%%%%%%%%%%%%%%%%%%
    $v_{\star}$ (\si{km.s^{-1}}) 
    & [$69,100$]                 
    & [$29,40$] 
    & (5)     \\
    %%%%%%%%%%%%%%%%%%%%%
    $\dot{M}$ (\si{M_\odot.yr^{-1}}) 
    & \num{9e-6}                        
    & \num{1.3e-6}                      
    & (8),(9) \\
    %%%%%%%%%%%%%%%%%%%%%
    $v_\infty$ (\si{km.s^{-1}}) 
    & 2300                       
    & 2000                       
    & (4),(9) \\
    %%%%%%%%%%%%%%%%%%%%%
    $n_\mathrm{H}$ (cm$^{-3}$) 
    & 15                       
    & 50                       
    & (8),(7) \\
    %%%%%%%%%%%%%%%%%%%%%
    $R_\mathrm{SO}$ (\arcmin) 
    & 2.72              
    & 0.42              
    & (5) \\
    %%%%%%%%%%%%%%%%%%%%%
    \bottomrule
    %%%%%%%%%%%%%%%%%%%%%
    
\end{tabular}
\begin{tablenotes}
      \small
      \item \textbf{Notes.} $R_\mathrm{SO}$ is the calculated standoff distance of the bow shock for the given parameters, $d$, $v_\star$, $v_\infty$, $\dot{M}$ and $n_\mathrm{H}$.
      \item \textbf{References.} 
    (1) \citetads{2021A&A...649A...1G},
    (2) \citetads{2016ApJS..224....4M}, 
    (3) \citetads{2011ApJS..193...24S},
    (4) \citetads{2007A&A...467L..23C}, 
    (5) this work,
    (6) \citetads{2010ApJ...710..549K},\citetads{2018A&A...616A...1G},
    (7) \citetads{2019A&A...625A...4G},
    (8) \citetads{2020MNRAS.495.3041T}, 
    (9) \citetads{2018A&A...617A..13D} 
    \end{tablenotes}
  \end{threeparttable}
\end{table}
%%%%%%%%%%%%%%%%%%%%%%%%%%%%%%%%%%%%%%%
%%%%%%%%%%%%%%%%%%%%%%%%%%%%%%%%%%%%%%%

\begin{figure*}
    \centering
    \includegraphics[width=\linewidth, trim={0 1.5cm 0 1cm}, clip]{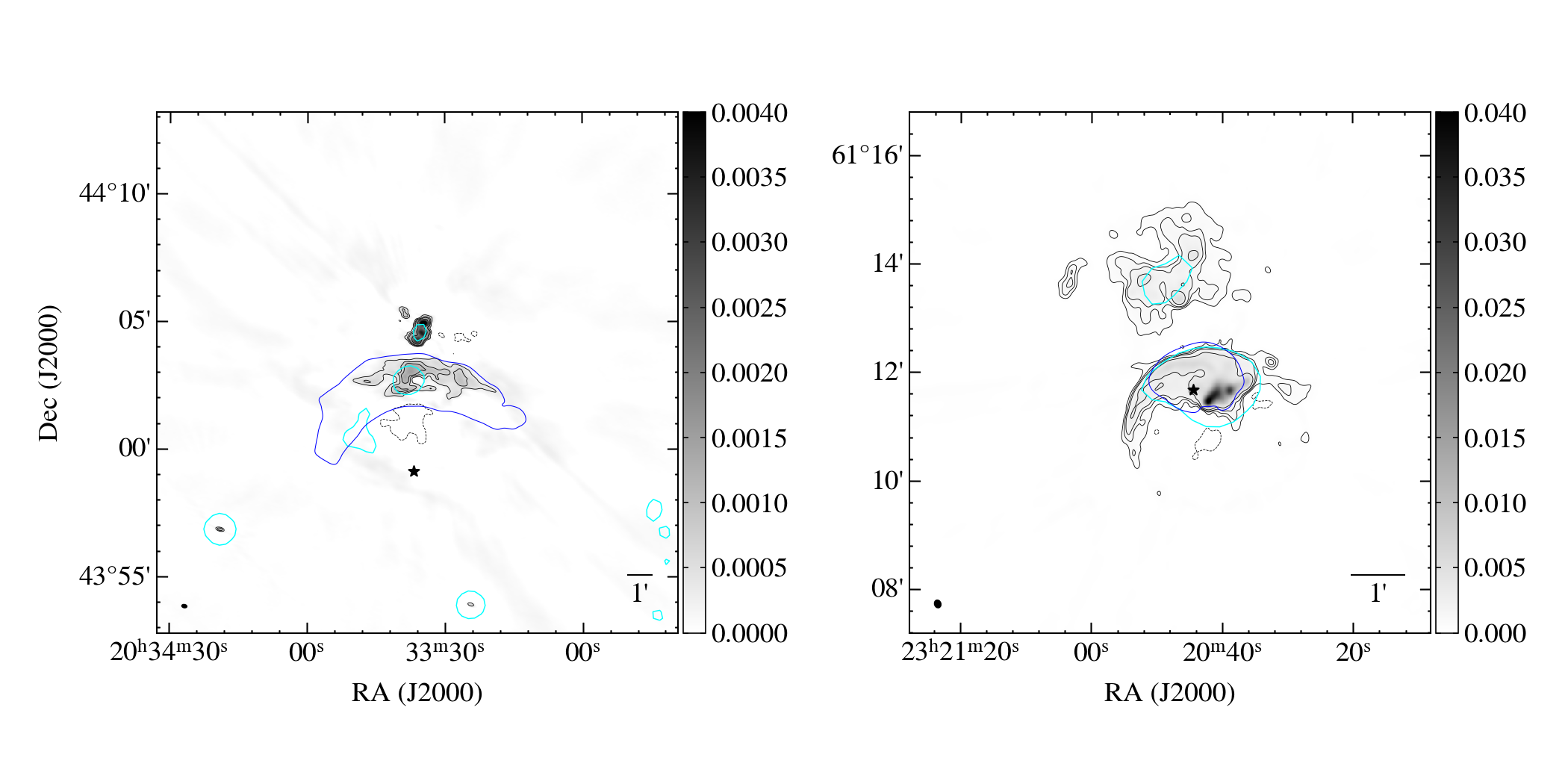}
    \caption{
    The (non primary beam corrected) images of BSBD43 (left) and BSBD60 (right) in radio with contours at \SIrange{4}{12}{GHz} (VLA, black), at \SI{1.4}{GHz} (NVSS, cyan, contour at \SI{20}{mJy}), and infrared at \SI{22}{\mu m} (WISE, blue). 
    The star is symbolised with an asterisk and the VLA beam size at the bottom left with a solid black ellipse. 
    The VLA contour levels are $-$5, 5, 10, 15, 30 times the rms of \SI{0.06}{mJy.beam^{-1}} for BSBD43, whereas the rms for BSBD60 is \SI{0.1}{mJy.beam^{-1}}. 
    The contours for NVSS and WISE were chosen in order to highlight the most interesting features of each image.
    The greyscale shows the intensity in \si{Jy.beam^{-1}}.}
    \label{fig:all_int}
\end{figure*}

%%%%%%%%%%%%%%%%%%%%%%%%%%%%%%%%%%%%%%%
\subsection{\texorpdfstring{\bdforty}{BSBD43}} \label{sub:tar_bd43}
%%%%%%%%%%%%%%%%%%%%%%%%%%%%%%%%%%%%%%%
\bdforty is an O4If blue supergiant runaway star originating from the Cygnus OB2 association (\citeads{2007A&A...467L..23C}, \citeads{2008A&A...490.1071G}), located at a distance of \SI{1.72}{kpc} \citepads{2021A&A...649A...1G}. 
With a stellar mass of $\sim$\SI{70}{M_\odot}, a velocity of \SIrange{69}{100}{km.s^{-1}} (see \ref{subsec:sp_vel}) and a wind speed of \SI{2300}{km.s^{-1}}, \bdforty is a very massive runaway star.
Its bow shock (hereafter referred to as BSBD43) is brightest in the infrared \citepads[e.g.][]{2016ApJS..227...18K}, but it is also visible at radio frequencies (\citeadsalias{2010A&A...517L..10B}, \citeadsalias{2021MNRAS.503.2514B}).
The latter have found the bow shock to emit both TE and NTE, which, so far, is the sole evidence of non-thermal processes in radio for such shocks.
Other studies have only obtained upper limits at higher energies \citepads[e.g.][]{2016ApJ...821...79T}.
From the brightness of the nebula and its size, the mean H number density of the \gls{ism}, $n_\mathrm{H}$, is estimated to be $\sim$\SI{15}{cm^{-3}} \citepads{2018A&A...617A..13D}.
The mass-loss rate, $\dot{M}$, of this star is uncertain, with values ranging from \SI{2.5e-6}{M_\odot.yr^{-1}} to \SI{2e-4}{M_\odot.yr^{-1}} \citepads{2016ApJ...821...79T}; we assume \SI{9e-6}{M_\odot.yr^{-1}} from \citetads{2018A&A...617A..13D}, appropriate for a star of its spectral type.

%%%%%%%%%%%%%%%%%%%%%%%%%%%%%%%%%%%%%%%
\subsection{\texorpdfstring{\bdsixty}{BSBD60}} \label{sub:tar_ngc}
%%%%%%%%%%%%%%%%%%%%%%%%%%%%%%%%%%%%%%%
\bdsixty is an O6.5(n)fp star \citepads{2011ApJS..193...24S}, surrounded by its wind-blown nebula, \ngc, commonly known as the Bubble Nebula \citepads{1995A&A...295..509C}.
The Bubble Nebula is prominent at infrared and optical wavelengths (\citeads{2002AJ....124.3313M}, \citeads{2019A&A...625A...4G}).
The high peculiar velocity of the star suggests that the nebula could be a bow shock (hereafter referred to as BSBD60).
In fact, simulations by \citetads{2019A&A...625A...4G} have shown that the bow shock model is capable of reproducing the optical and infrared observations, further supporting the hypothesis.
They have also suggested that similarities with BSBD43 make BSBD60 a good candidate for detecting both TE and NTE.

X-ray observations by \citetads{2020MNRAS.495.3041T} with \emph{XMM-Newton} detected \bdsixty but no extended X-ray emission (TE nor NTE).
The latter also estimate a mass of \SI{27}{M_\odot} based on a distance of \SI{2.5}{kpc}, but a newly measured parallax from \citetads{2021A&A...649A...1G} revised the distance to \SI{3.0}{kpc}, and therefore the calculated mass should be rescaled to \SI{39}{M_\odot}.

The only previous radio observations of \ngc are low-resolution CO maps by \citetads{1982MNRAS.201..429T}; the observations we present here are the first high-resolution continuum maps at~GHz frequencies.

\bdsixty appears to be less massive, less evolved, and has a weaker wind than \bdforty, and it is moving more slowly through the \gls{ism}.
Nevertheless, it is moving through a denser \gls{ism} than \bdforty (estimated n$_\mathrm{H}\sim$\SI{50}{cm^{-3}}, \citeads{2019A&A...625A...4G}), and this gives the nebula a large emission measure resulting in bright optical line emission and continuum radio emission.

\subsection{Space velocity}\label{subsec:sp_vel}
We follow the method in \citetads{2019A&A...625A...4G} to calculate the space velocity of the two stars using \emph{Gaia} EDR3 data \citepads{2021A&A...649A...1G}.
We take the distance to the Galactic Centre of \SI{8.0}{kpc}, the circular Galactic rotation velocity of \SI{240}{km.s^{-1}} \citepads{2009ApJ...705.1548R}, and for the Solar peculiar velocity $(U_\odot, V_\odot, W_\odot) = (11.1, 12.2, 7.3)$ \si{km.s^{-1}} \citepads{2010MNRAS.403.1829S}.
Using these we obtain the peculiar velocity in the plane of the sky, $v_\mathrm{tr}$ and the peculiar radial velocity $v_\mathrm{r}$.
For \bdforty we obtain $v_\mathrm{r} = \mathrm{RV}+12.5\pm0.1$\,\si{km.s^{-1}} and for \bdsixty $v_\mathrm{r} = \mathrm{RV}+42\pm2$\,\si{km.s^{-1}}, where RV is the heliocentric radial velocity with values taken from the references cited in the table.
For both stars there are varying estimates of the radial velocity, with extreme differences between \citetads{2010ApJ...710..549K} and \citetads{2018A&A...616A...1G} for the radial velocity of \bdforty.
This results in significant uncertainty in $v_\mathrm{r}$ and the total peculiar velocity, $v_\star = \sqrt{v_\mathrm{tr}^2+v_\mathrm{r}^2}$.

%%%%%%%%%%%%%%%%%%%%%%%%%%%%%%%%%%%%%%%
% 03 - RADIO OBSERVATIONS
%%%%%%%%%%%%%%%%%%%%%%%%%%%%%%%%%%%%%%%
\section{Radio Observations}\label{sec:obs}
\begin{figure*}

    \centering
    \includegraphics[width=0.43\linewidth,
    ]{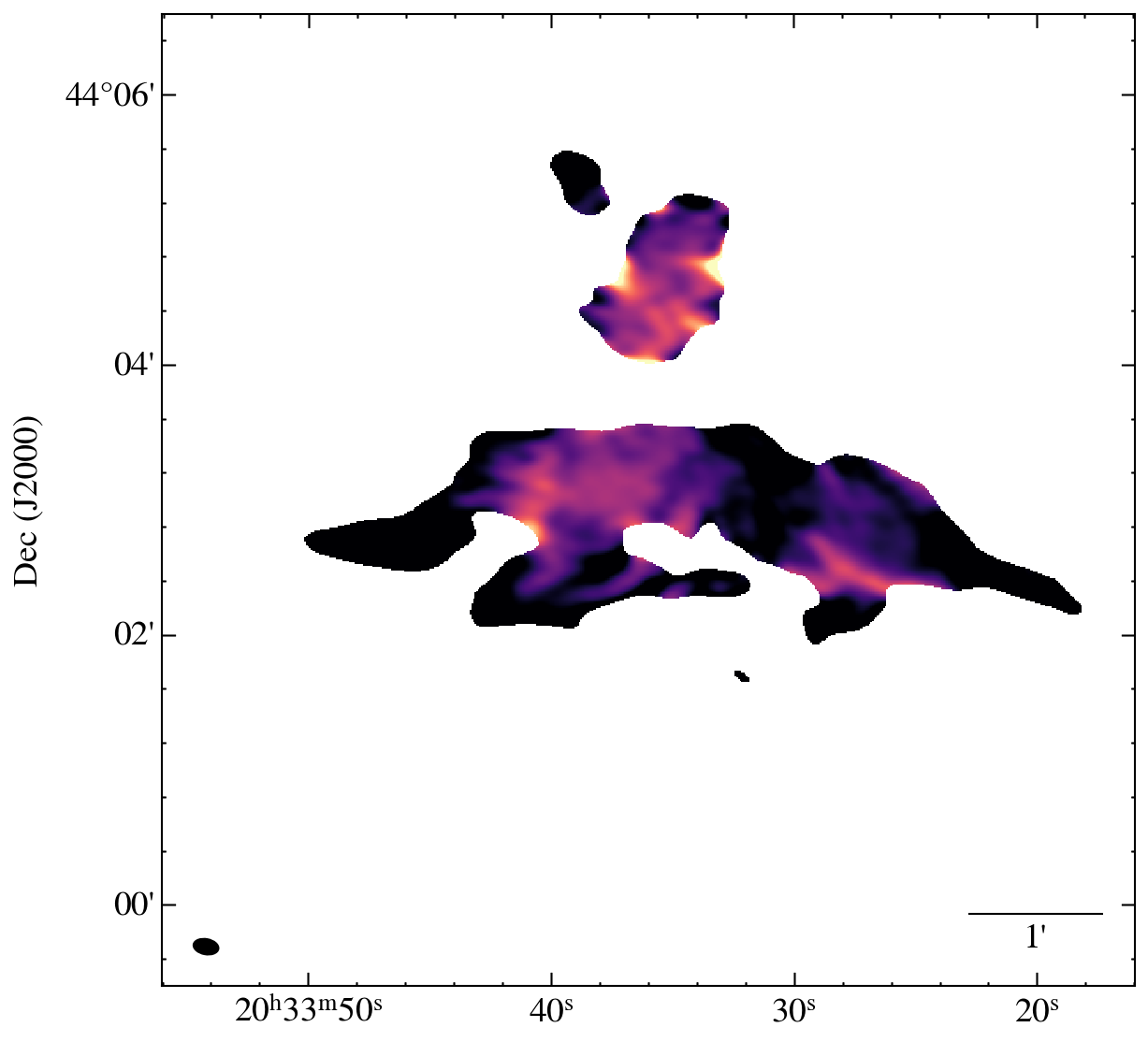}
     \includegraphics[width=0.45\linewidth, 
     ]{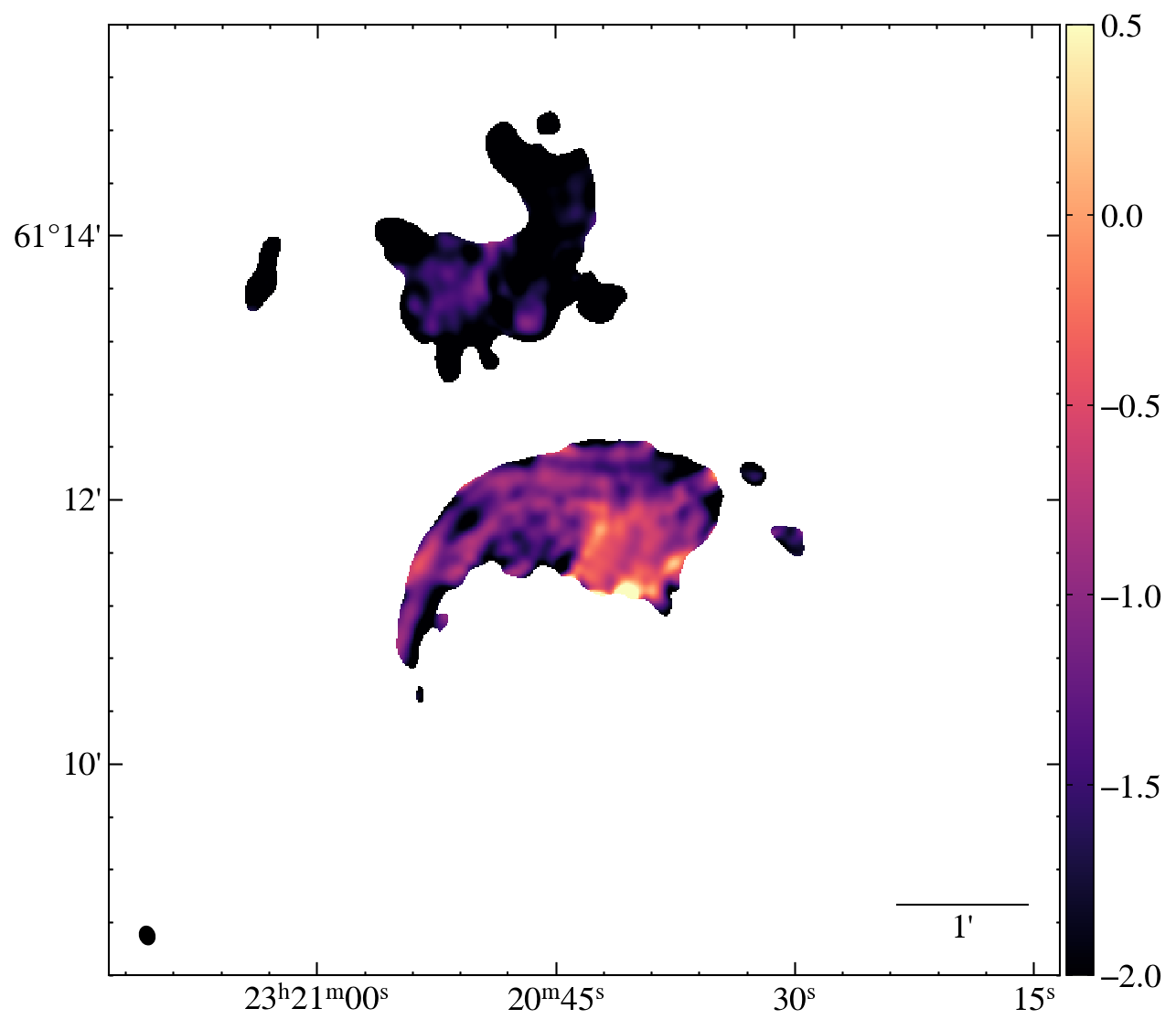}
    \\[\smallskipamount]
   \includegraphics[width=0.435\linewidth, 
   ]{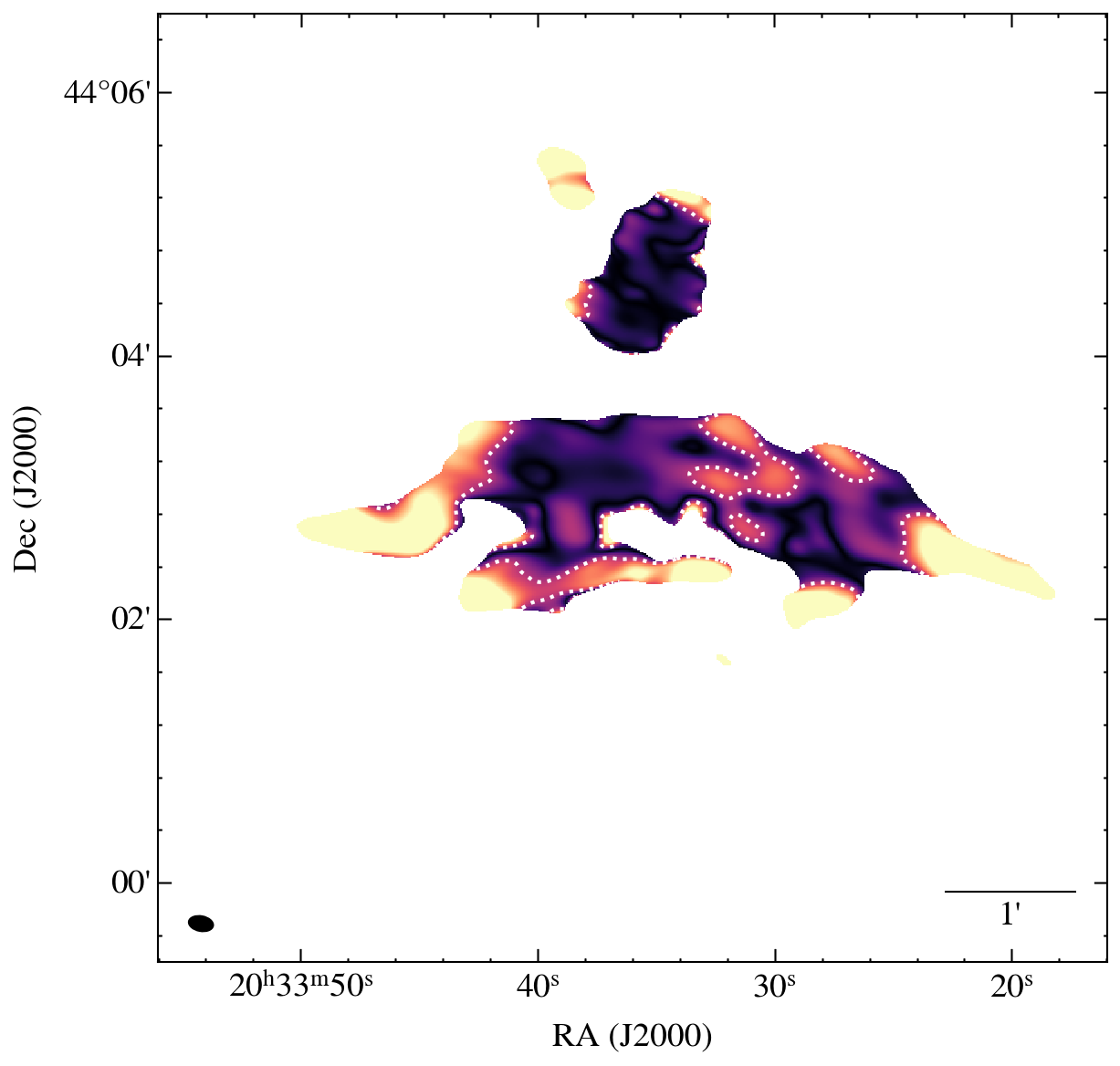}
    \includegraphics[width=0.45\linewidth, 
    ]{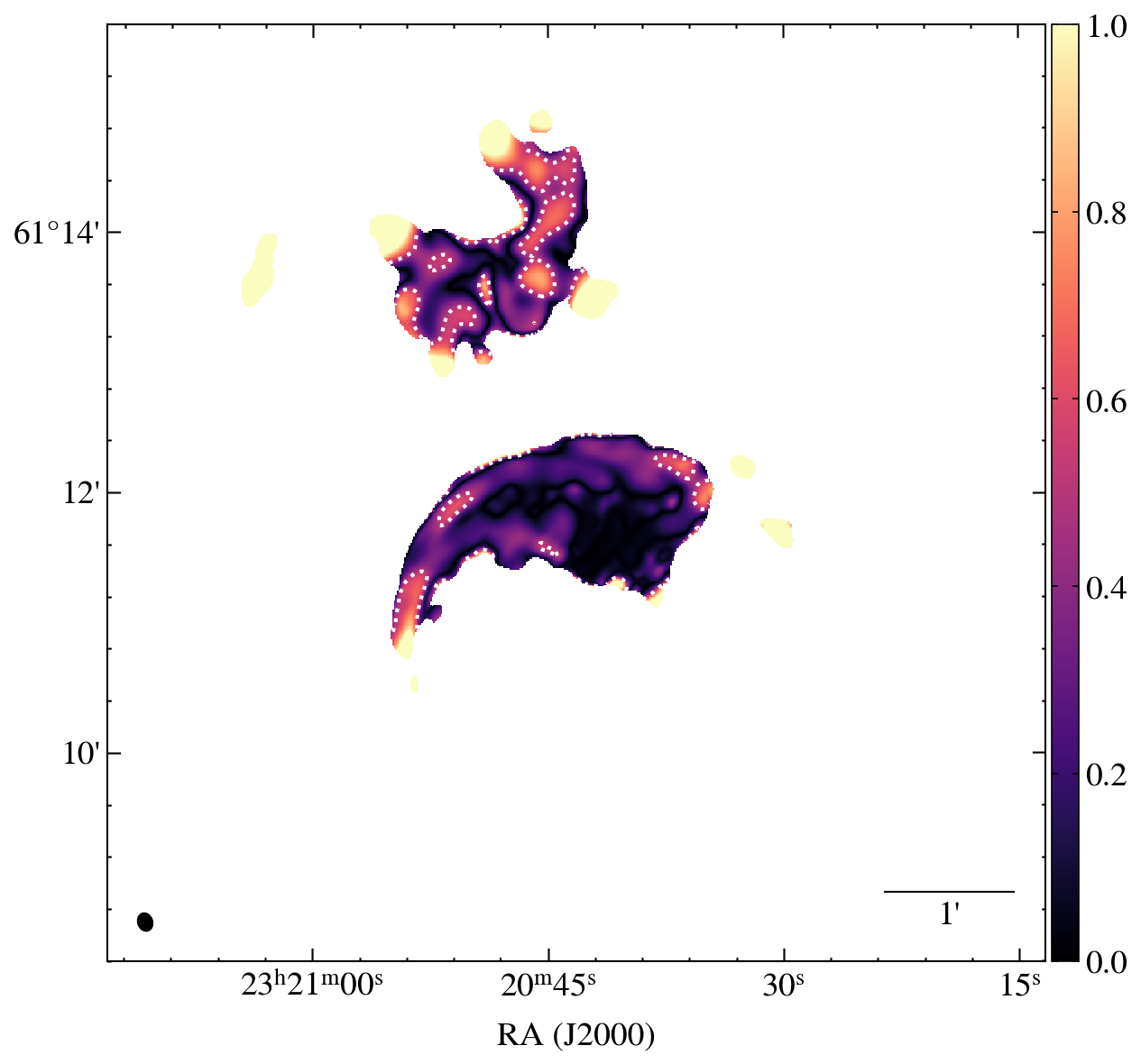}
    \caption{Maps of the Spectral Index $\alpha$ (top) and its uncertainty $\alpha_\mathrm{err}$ (bottom) for BSBD43 (left) and BSBD60 (right). 
    Only pixels with significance above 5$\sigma$ were selected, which is \SI{0.3}{mJy.beam^{-1}} for BSBD43 and \SI{1}{mJy.beam^{-1}} for BSBD60, while white dotted contours surround errors equal to 0.5. 
    The maps are scaled for $\alpha$=[$-2,0.5$] and $\alpha_{\mathrm{err}}$=[0,1]. 
    All maps are primary beam corrected.}
    \label{fig:bd_spindx_all}
\end{figure*}

\subsection{Interferometric Data}
In order to map the radio continuum of BSBD43 and BSBD60, observations were carried out with the VLA in New Mexico, on 16$^\mathrm{th}$ and 19$^\mathrm{th}$ November 2019 (Project ID, PID: 19B-105), covering the frequencies of \SIrange{4}{12}{\GHz}.
The mobile 27 antennas of the VLA are in a Y-shaped arrangement, while their distance can be modified to provide various combinations of angular resolution and surface brightness sensitivity in all the available frequencies.
We chose the tightest configuration, D, with antennae separation from B$_\mathrm{min}$ \SI{35}{m} to B$_\mathrm{max}$ \SI{1}{km}, observed at the C and X band (\SIrange{4}{8}{\GHz} and \SIrange{8}{12}{\GHz} respectively) for optimal signal to noise ratio for such extended sources.
The bands consisted of a number of \SI{128}{MHz} wide spectral windows (see Sect. \ref{subsub:ifrd}).
Both targets were observed once for each band for $\sim$\SI{30}{\min}.
For the flux density scale and bandpass calibration, 3C\,48 was used, while for the phase and amplitude calibration, J2007+4029 and J2230+6946 were observed before and after each scan of BSBD43 and BSBD60 correspondingly.
The raw data acquired from the observations were reduced using the VLA calibration pipeline included in the \textit{Common Astronomy Software Applications}  \citepads[CASA,][]{2007ASPC..376..127M}.
We used the CASA version 5.7 (6.1 for the pipeline) and the National Radio Astronomy Obervatory's (NRAO)\footnote{The National Radio Astronomy Observatory is a facility of the National Science Foundation operated under cooperative agreement by Associated Universities, Inc.} facilities in order to reduce our data.

\subsection{Single-Dish Data}
There are two main difficulties when trying to measure extended emission accurately using an interferometer: the short-spacing problem (SSP) and the finite size of the primary beam.
For single-pointing observations like ours, the sensitivity of the measurement decreases with angular distance from the centre of the map because of the primary beam response.
We can correct for this by scaling the map, but this also increases the noise level far from the map centre.
What is more, beyond the first zero of the primary beam we cannot get any further useful information.
The latter, since the primary beam size depends directly on the observed frequency, affects our observations at high frequencies for BSBD43, but less so for BSBD60 because it has a smaller angular extent.
The SSP, on the other hand, arises because the antennas have a finite number of baselines and some short baselines are inaccessible because of the antenna layout.
This can be potentially rectified by combining interferometric data with single dish observations \citepads[e.g.][]{2019AJ....158....3R} to help fill the $uv$-plane.

During September 2020 we observed the BSBD43 and BSBD60 with the S45 broad-band receiver and the SPEctro-POLarimeter (SPECPOL) backend of the Effelsberg 100 meter radio telescope\footnote{The 100-m telescope at Effelsberg is operated by the Max-Planck-Institut für Radioastronomie (MPIFR) on behalf of the Max-Planck Gesellschaft (MPG).} 
at \SIrange{4}{8}{\GHz} (PID: 86-20), reaching a sensitivity of $\sim$\SI{3}{mJy.beam^{-1}}. 
At the beginning of each observing session, pointing and focus were checked and corrected by performing cross scans of 3C\,286. 
The same target was also adopted as our flux calibrator, with flux density of $\sim$\SI{7.3}{Jy} at \SI{5}{GHz} \citepads[see Eq.~(2) in][]{2021A&A...651A..85B}. 
Each image was scanned in two orthogonal directions, and the ``basket-weaving'' technique was used to reduce the scanning effects in the final image. 
For the lengths of the scans, they were designed to be \ang{;20;} for both BSBD43 and BSBD60.
The overhead dumps beyond the designed \ang{;20;} scan are discarded in the image restoration, because they are not well sampled. 
The analysis was preformed with the NOD3 software package \citepads{2017A&A...606A..41M} where the radio frequency interference (RFI) affecting certain frequency ranges is ``flagged'' as part of the data editing process. 
The beam size is about \ang{;;140} at \SI{5}{GHz}. 

%%%%%%%%%%%%%%%%%%%%%%%%%%%%%%%%%%%%%%%
% 04 - RADIO EMISSION ANALYSIS
%%%%%%%%%%%%%%%%%%%%%%%%%%%%%%%%%%%%%%%
\section{Radio Emission Analysis \& Results}\label{sec:rea}
Following the imaging of BSBD43 and BSBD60, we estimated the spectral index of the bow shocks with several methods, both for our interferometric and single dish data, in order to determine the nature of their emission throughout the frequencies. 
In this section, we present our analysis.

%%%%%%%%%%%%%%%%%%%%%%%%%%%%%%%%%%%%%%%%
%%%%%%%%%%%%%%%%%%%%%%%%%%%%%%%%%%%%%%%%
\subsection{Imaging and Spectral Index Maps}\label{subs:im_and_spx}
%%%%%%%%%%%%%%%%%%%%%%%%%%%%%%%%%%%%%%%%
%%%%%%%%%%%%%%%%%%%%%%%%%%%%%%%%%%%%%%%%

\begin{figure*}
    \centering
    \includegraphics[width=0.33\linewidth,]{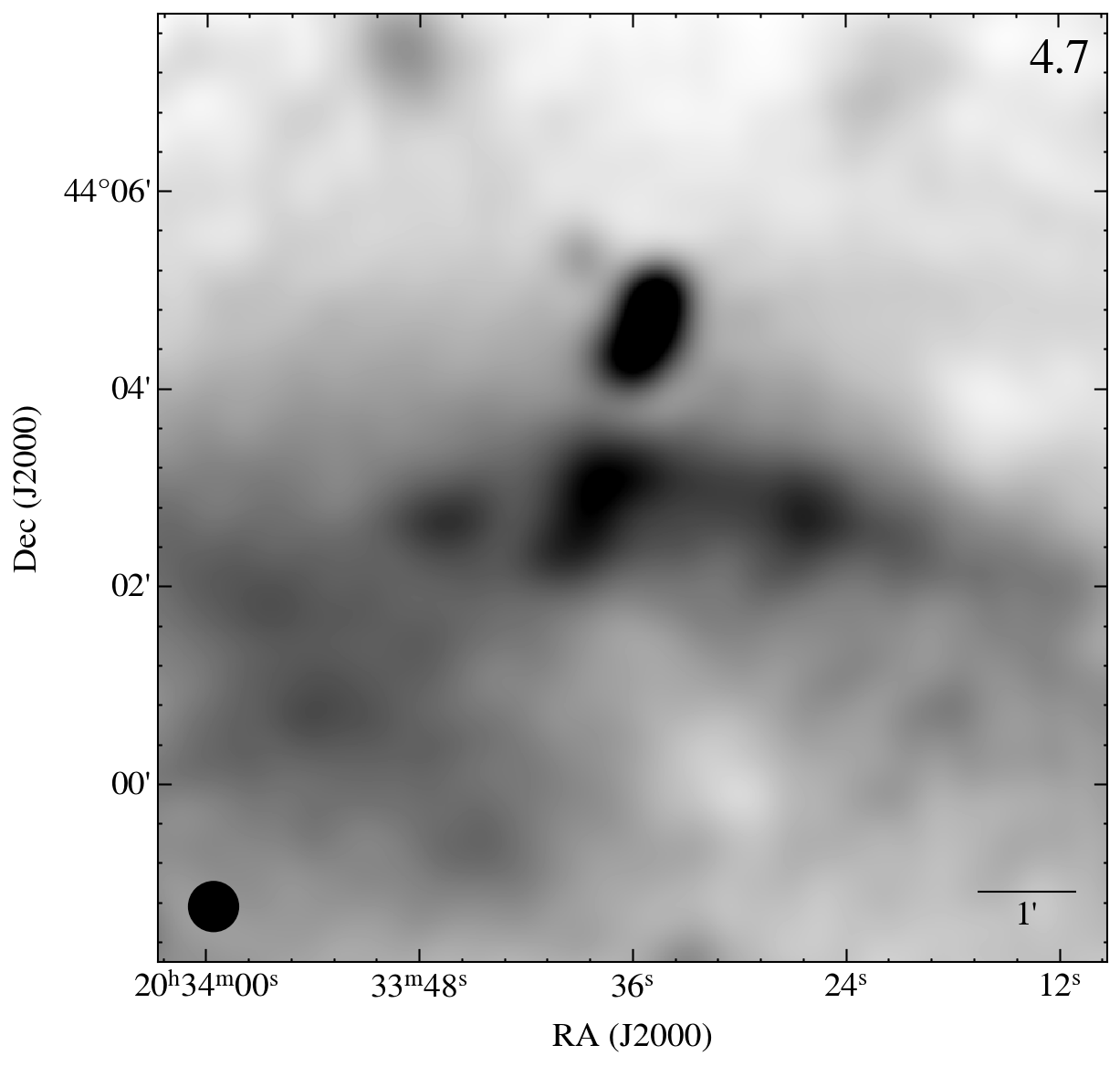}\hfill
    \includegraphics[width=0.33\linewidth,]{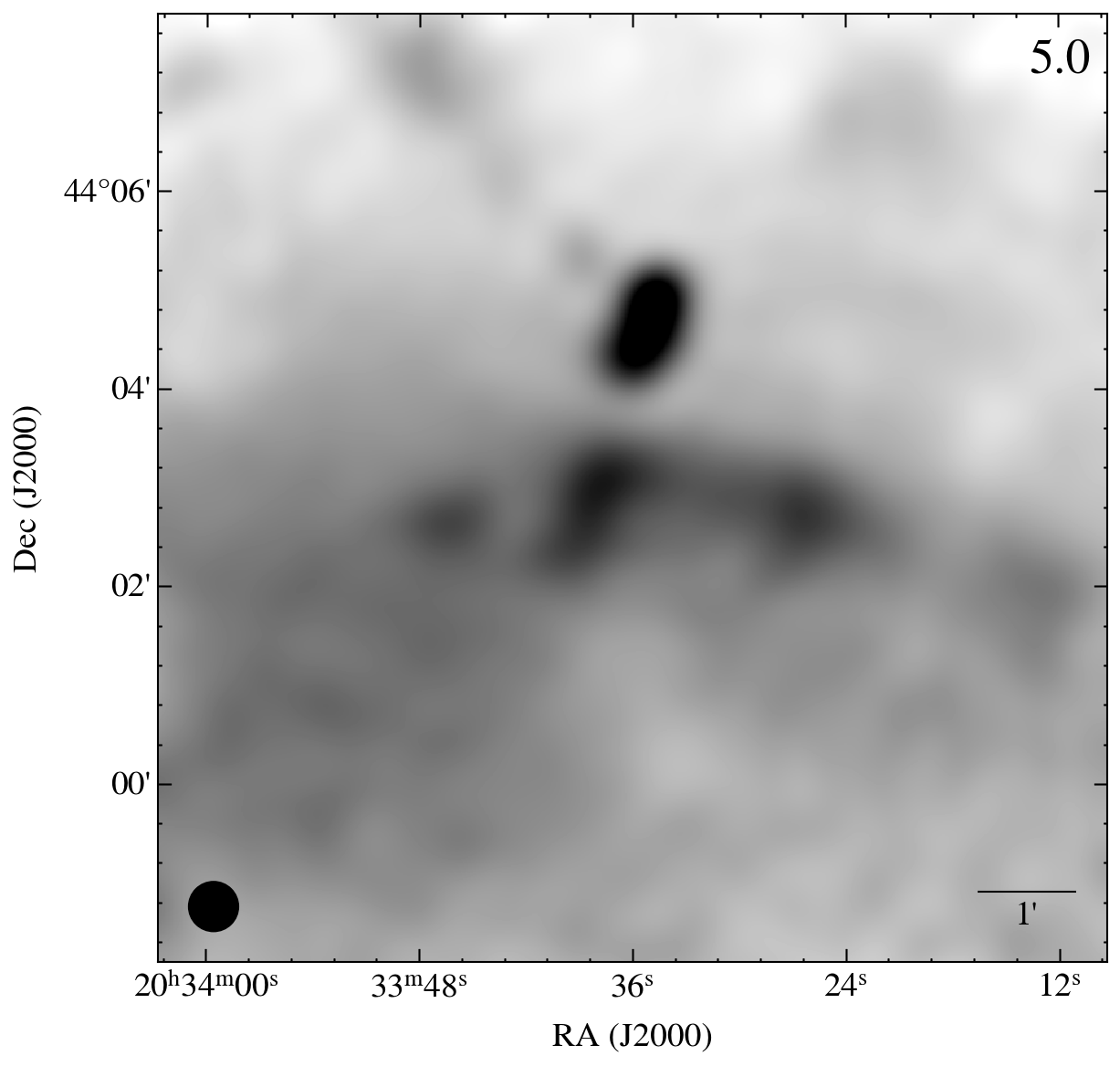}\hfill
    \includegraphics[width=0.33\linewidth,]{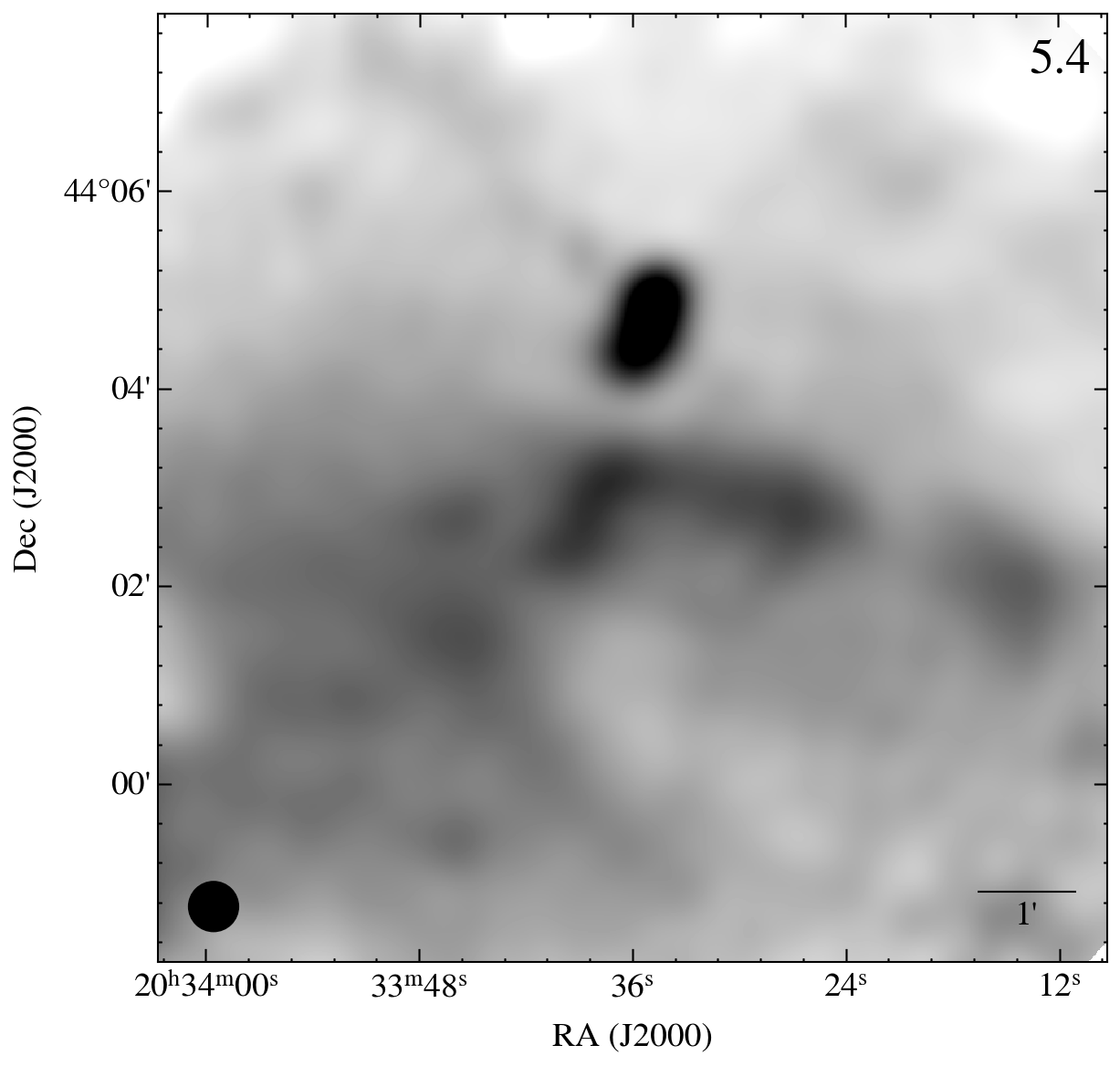}
    \\[\smallskipamount]
    \includegraphics[width=0.33\linewidth,]{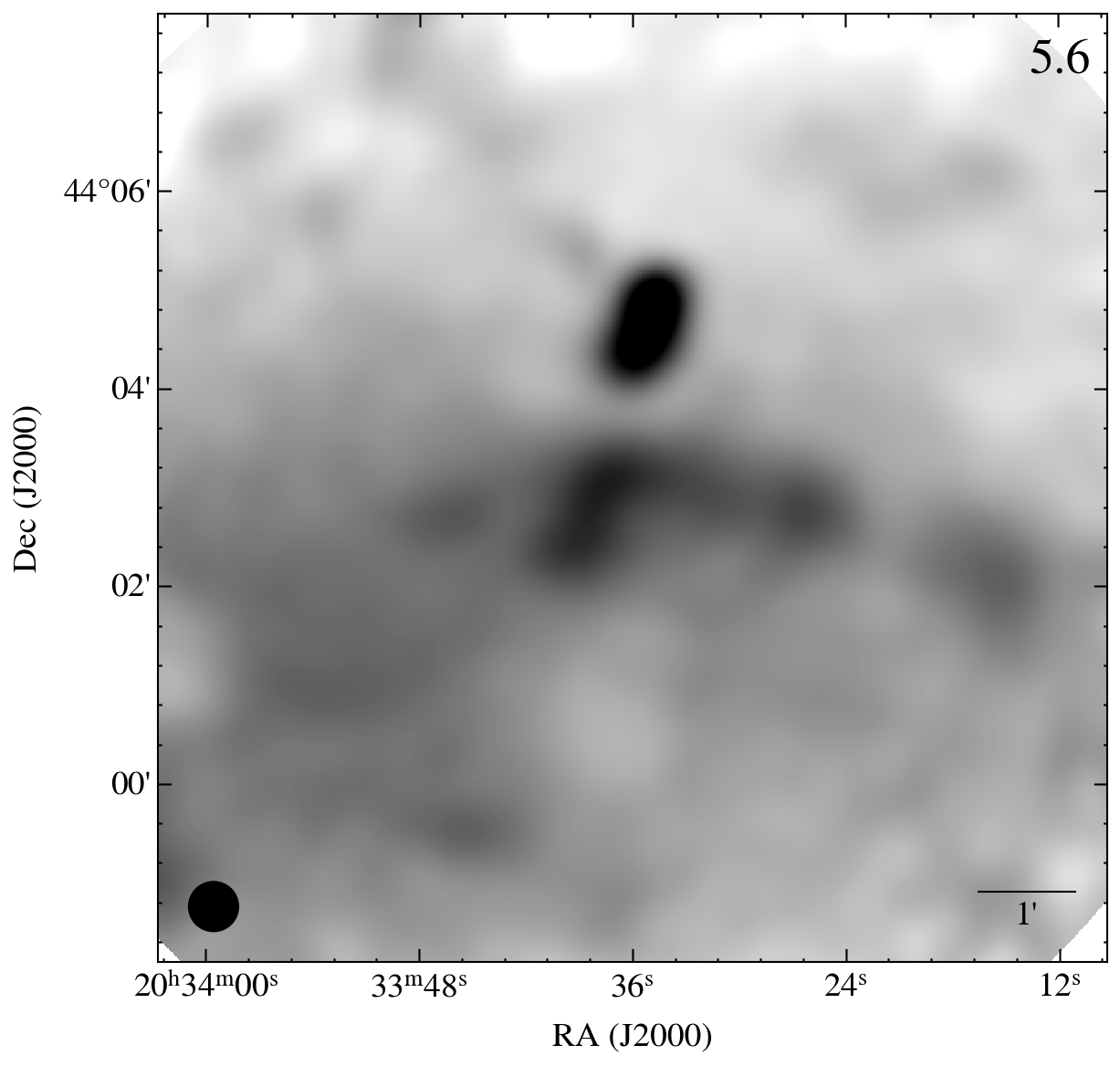}\hfill
    \includegraphics[width=0.33\linewidth,]{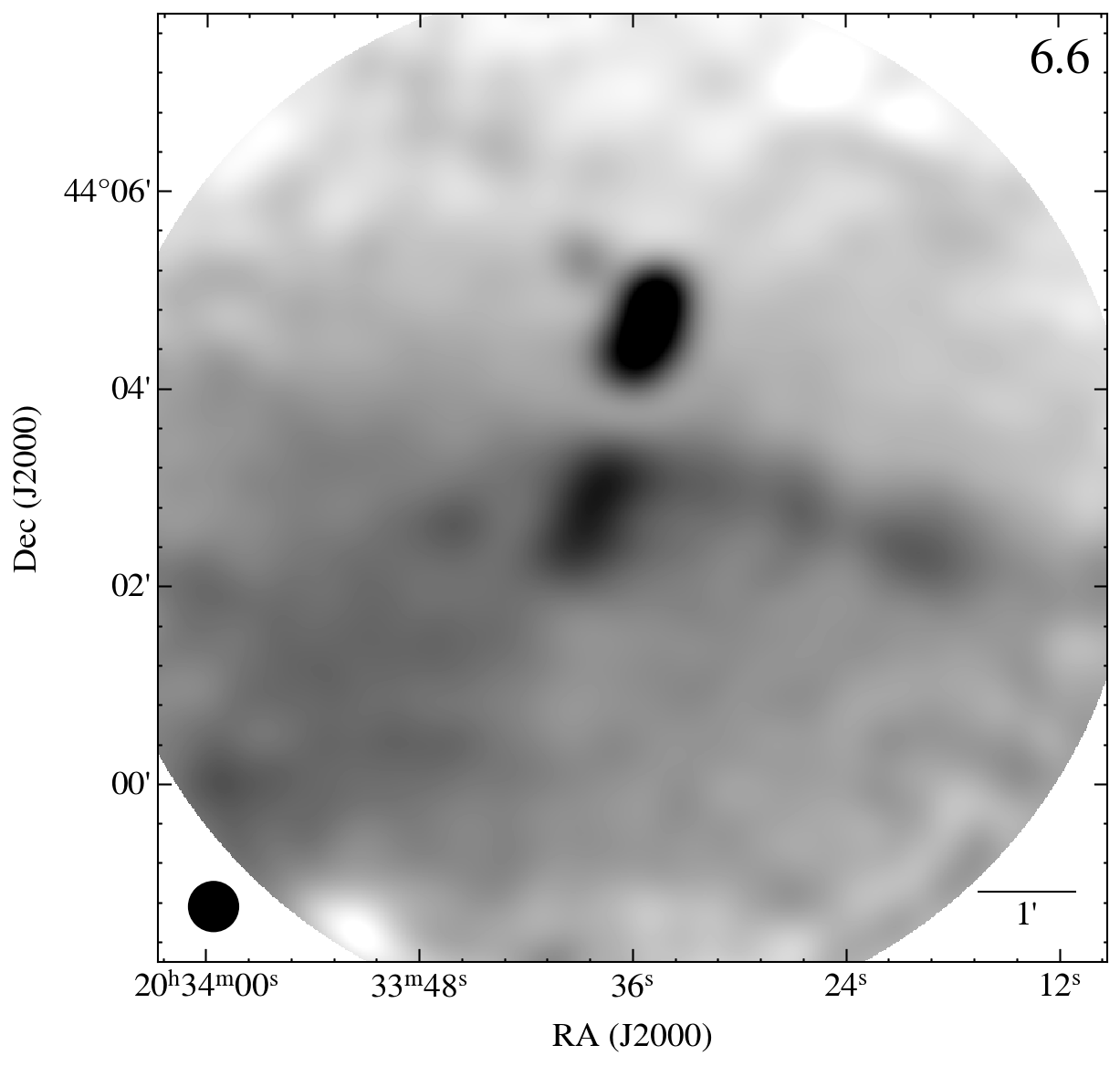}\hfill
    \includegraphics[width=0.33\linewidth,]{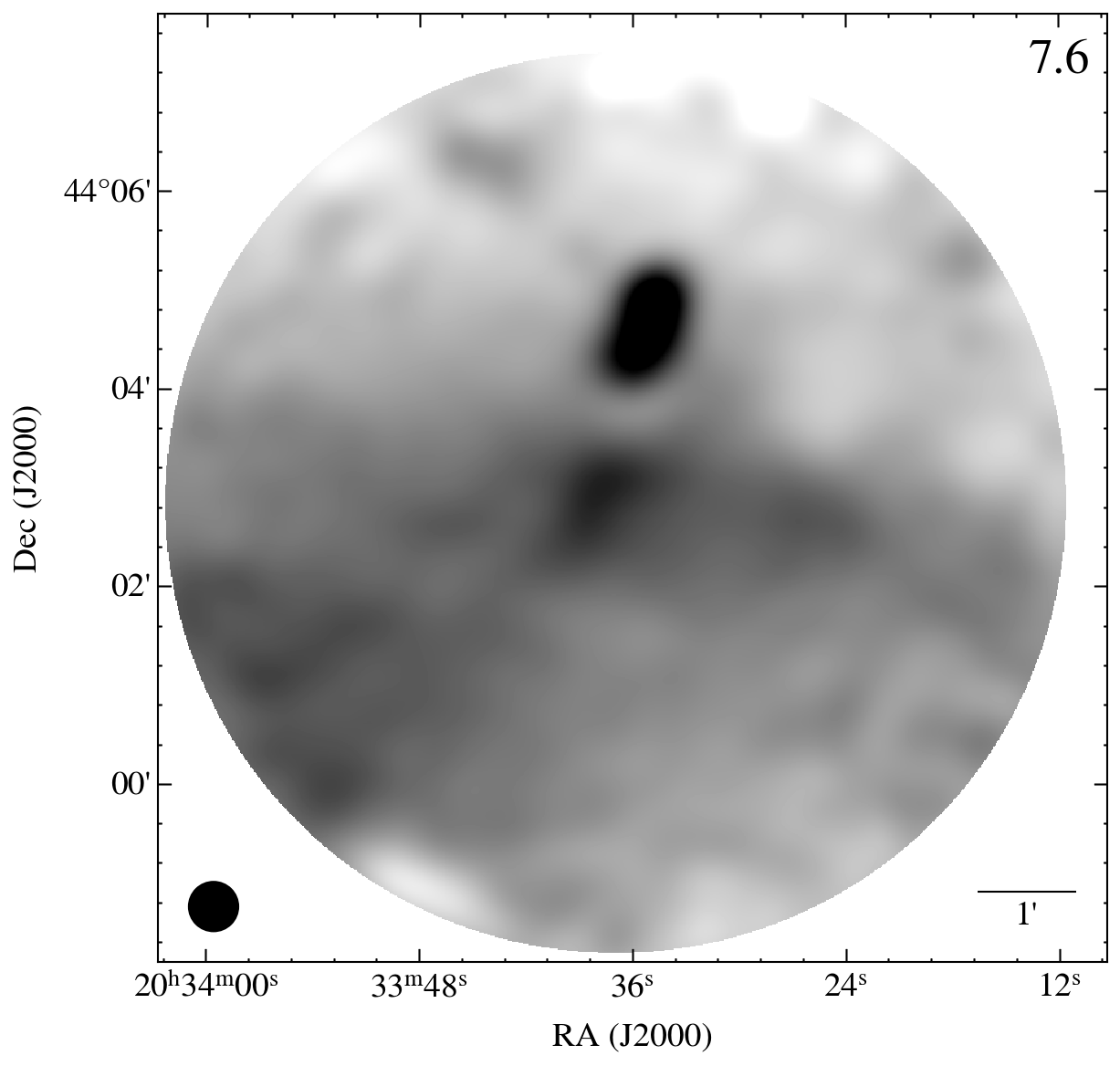}
    
    \caption{Feathered results for BSBD43. All images appear within the limits of [0,1.2] \si{MJy.sr^{-1}}. The central frequency of each image is given on the top right corner, in GHz.}
    \label{fig:comb_all_bd43}
\end{figure*}

\begin{figure*} %trim={<left> <lower> <right> <upper>}
    \centering
    \includegraphics[width=0.33\linewidth]{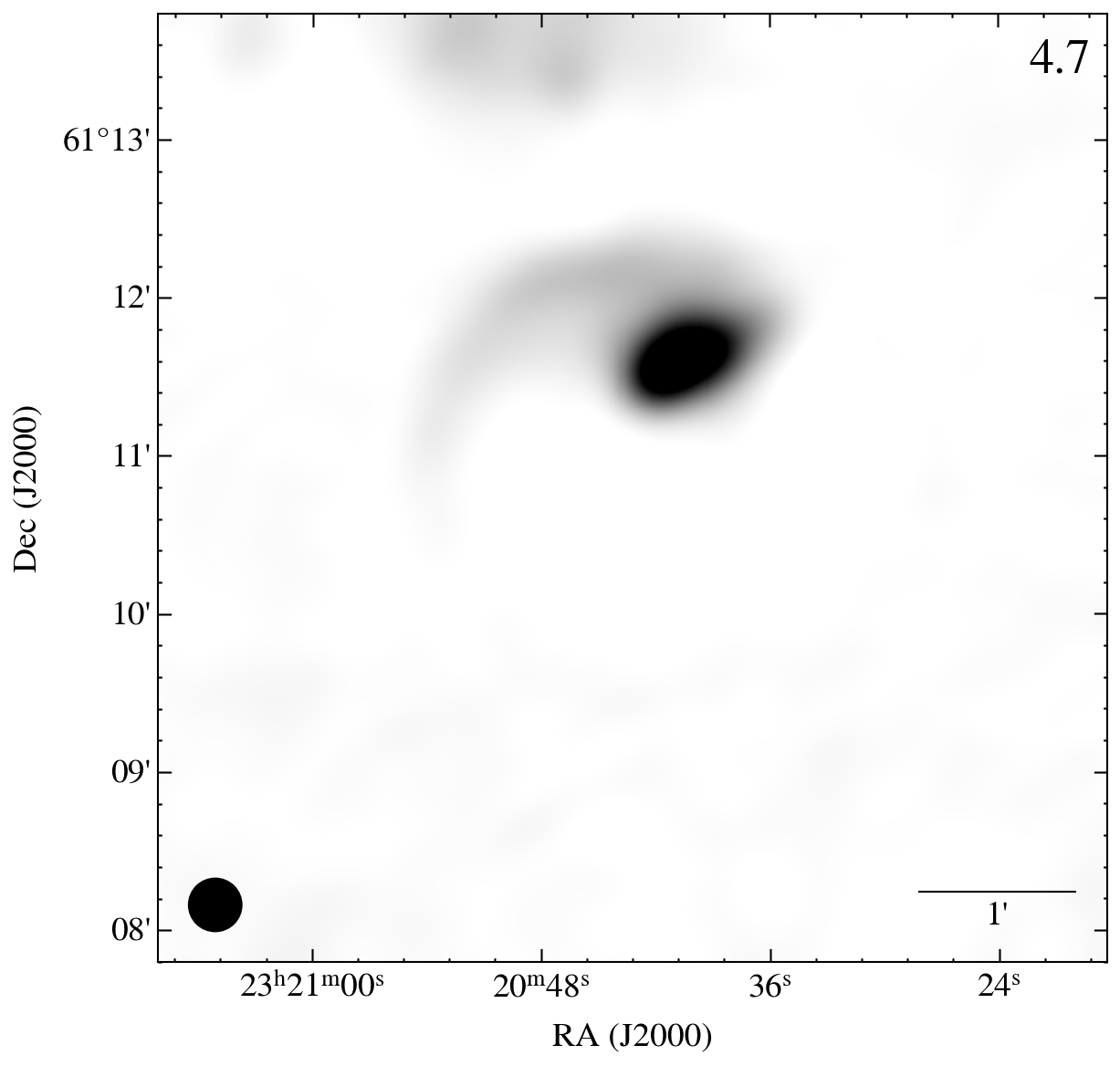}\hfill
    \includegraphics[width=0.33\linewidth,]{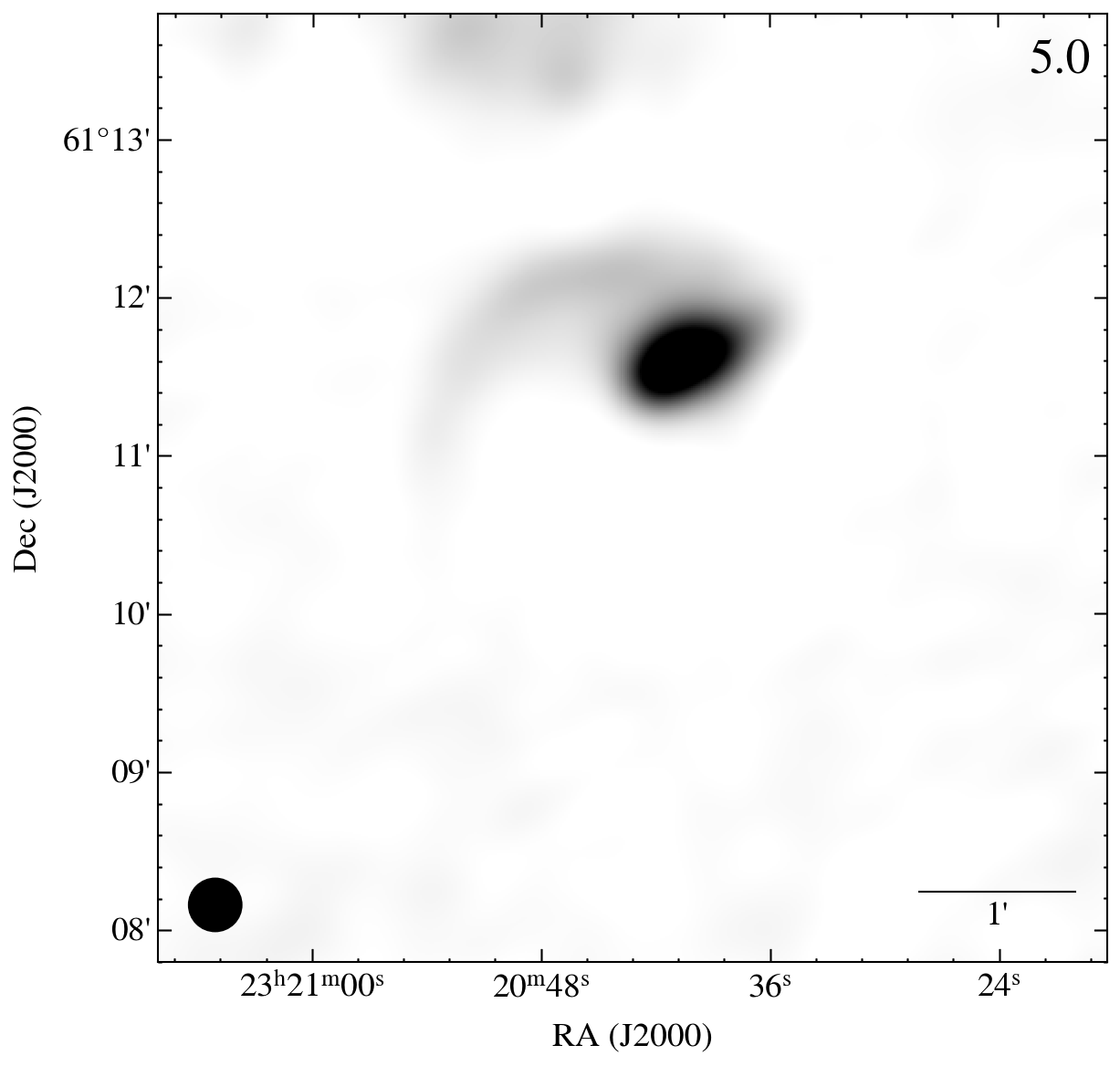}\hfill
    \includegraphics[width=0.33\linewidth,]{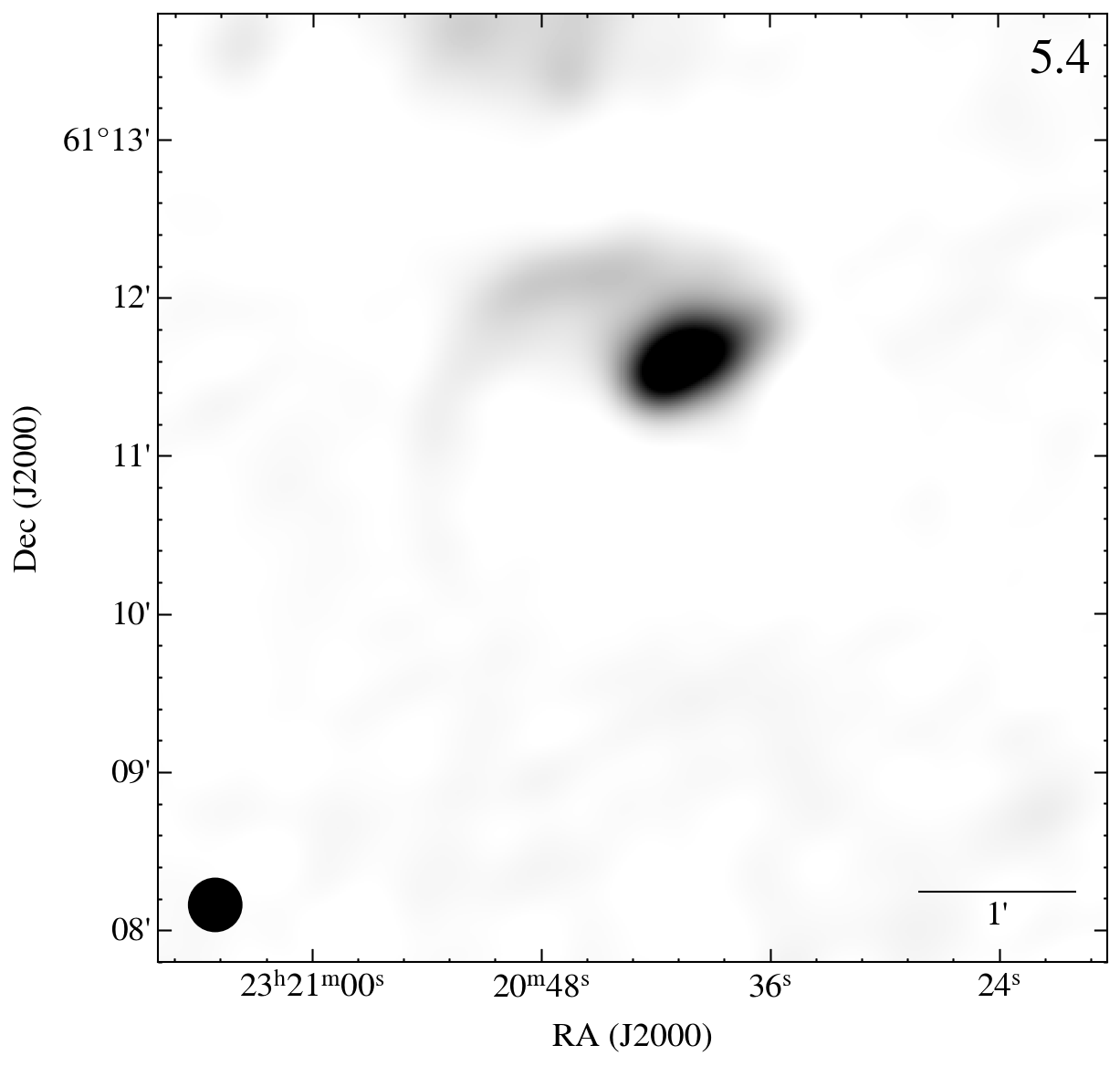}
    \\[\smallskipamount]
    \includegraphics[width=0.33\linewidth,]{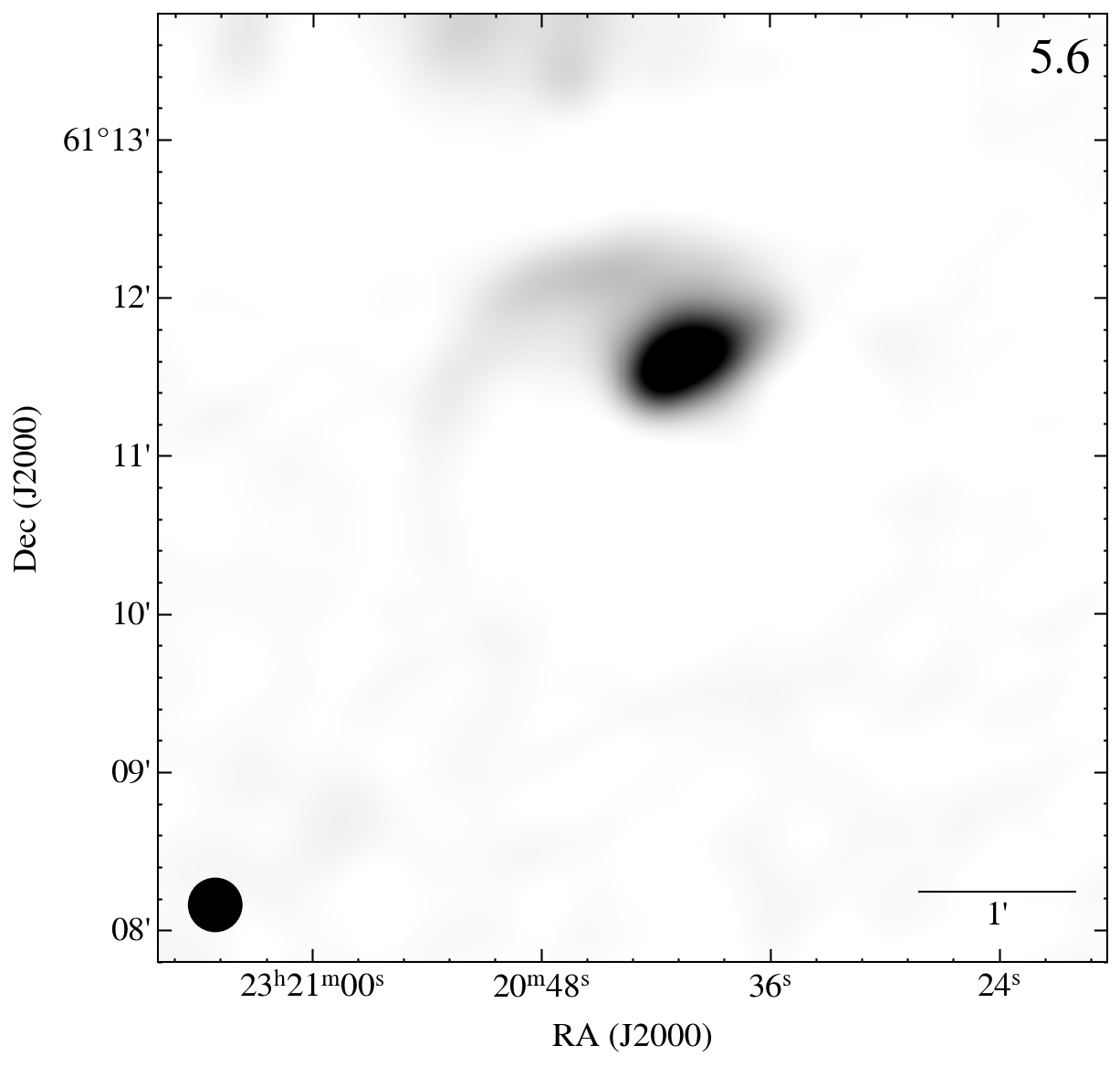}
    \includegraphics[width=0.33\linewidth,]{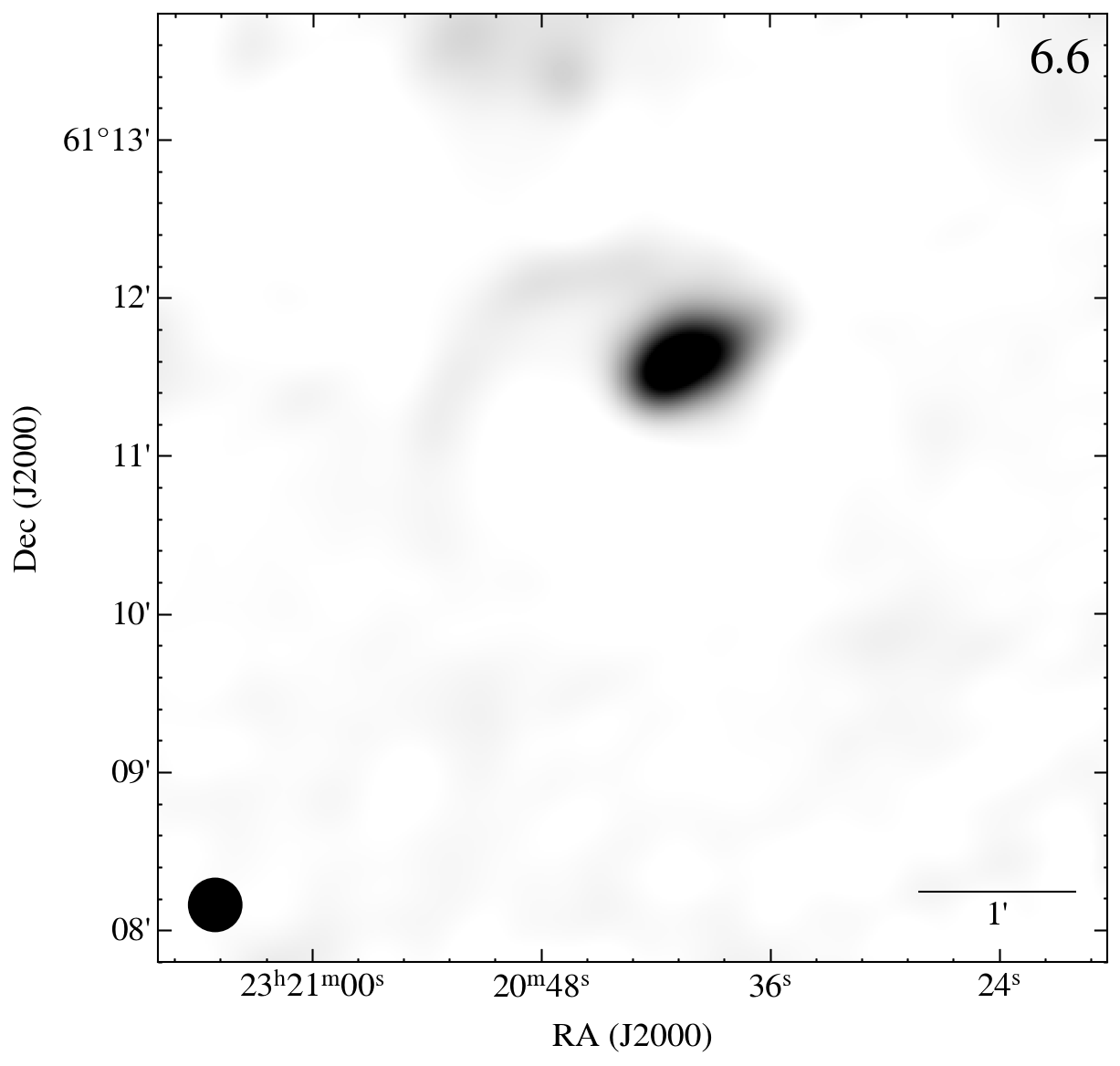}\hfill

    \caption{Same with Fig. \ref{fig:comb_all_bd43} but for BSBD60 only.  All images appear within the limits of [0,10] \si{MJy.sr^{-1}} for BSBD60.}
    \label{fig:comb_all_bd60}
\end{figure*}

\begin{figure*}
    \centering
    \includegraphics[width=1\columnwidth]{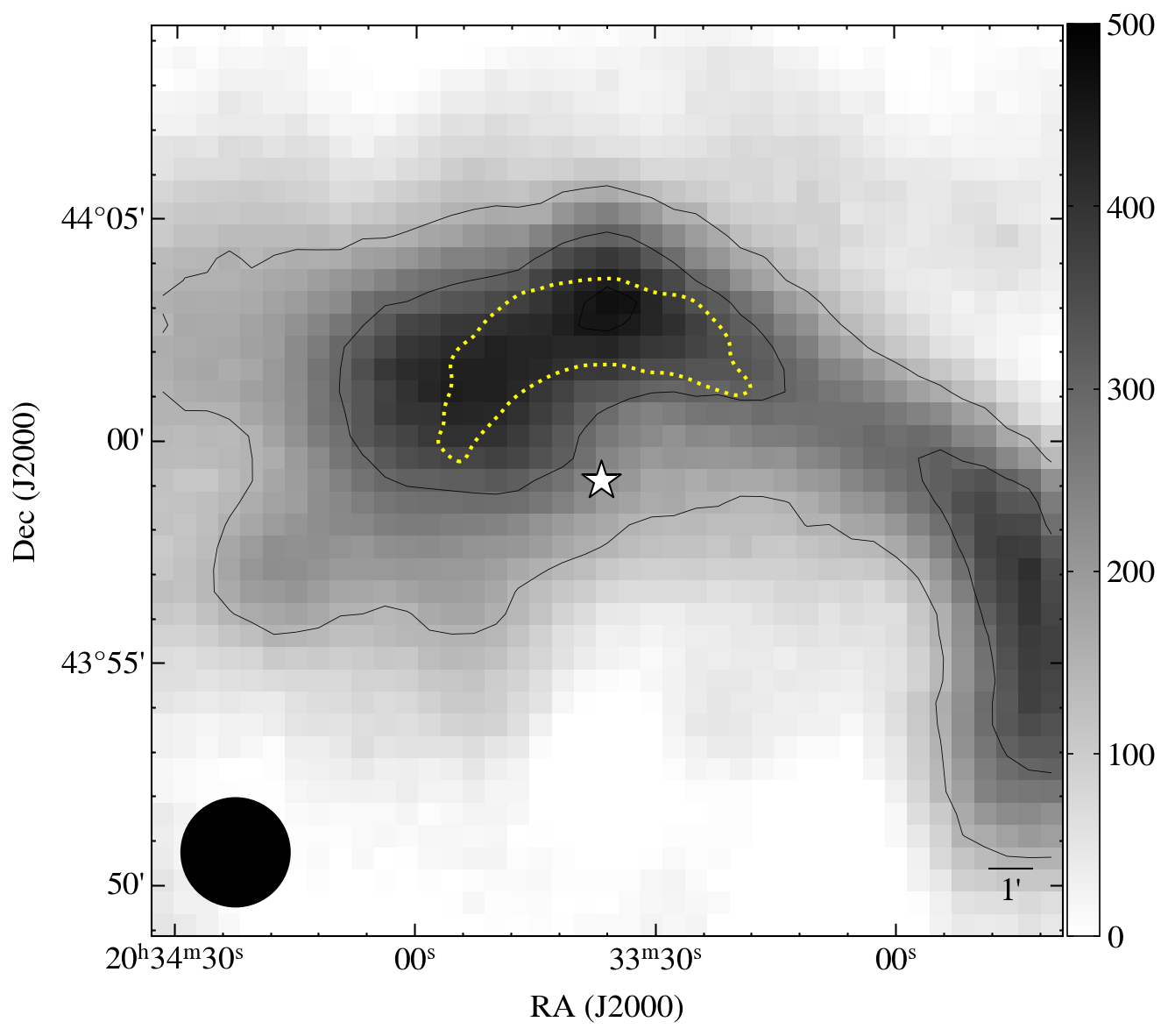}\hfill
    \includegraphics[width=1\columnwidth]{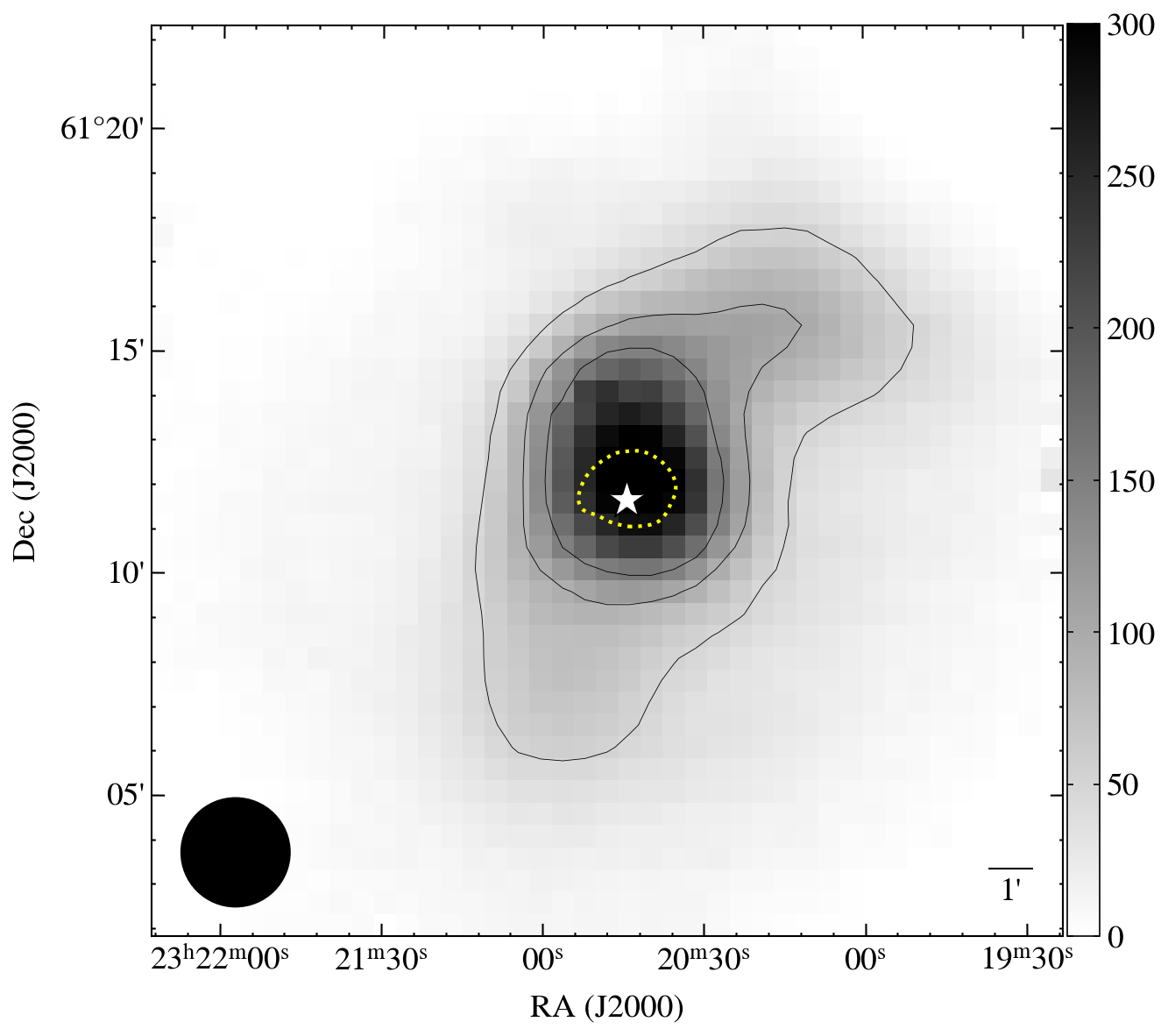}
    \caption{Intensity map of the Effelsberg data at \SI{4.5}{GHz}, BSBD43 left and BSBD60 on the right. 
    The beam size is represented as a solid black disc on the bottom left, while the star is represented with a white asterisk. 
    The greyscale shows the intensity in~\si{mJy.beam^{-1}}. 
    The contours represent emission of 5, 10 and 15 $\sigma$, which $\sigma$ is \SI{30}{mJy.beam^{-1}} and \SI{10}{mJy.beam^{-1}} for BSBD43 and BSBD60 correspondingly.
    The yellow dotted contours represent WISE emission at \SI{22}{\mu m}, of \SI{250}{dn} and \SI{580}{dn}.}
    \label{fig:eff_int}
\end{figure*}

\subsubsection{Interferometric Data}\label{subsub:ifrd}
%%%%%%%%%%%%%%%%%%%%%%%%%%%%%%%%%%%%%%%%%%%%%%%%%%%%%%%%%%%%%%%%%%%%%%%%%%%%%%%%
Firstly, we imaged our targets for the whole frequency range of \SIrange{4}{12}{GHz}, by combining all 64 spectral windows (SPWs) with a bandwidth of \SI{128}{MHz} each (Fig.~\ref{fig:all_int}).
The CASA task we used to do so, \emph{tclean}, contains several image reconstruction algorithms, such as CLEAN \citepads{1974A&AS...15..417H}, multi-scale \citepads[e.g.][]{2008ISTSP...2..793C}, as well as multi-scale multi-frequency  deconvolution algorithms \citepads[MS-MFS,][]{2011A&A...532A..71R}.
Since BSBD43 and BSBD60 were both observed in multiple SPWs, the frequency dependence of the synthesised beam has to be accounted for, as well as the fact that the sources are very extended.
With MS-MFS both were taken into account, leading us to choose it for our imaging process.

The final images (Fig.~\ref{fig:all_int}) were both cleaned for a diameter of $4\times\mathrm{FWHM}_\mathrm{PB}$, where $\mathrm{FWHM}_\mathrm{PB}$ is the Full Width Half Maximum (FWHM) of the primary beam (PB) of the instrument and it is estimated:
\begin{equation} \label{eq:fwhm_pb}
    \mathrm{FWHM}_\mathrm{PB}\,(\arcmin) \approx 42 \times \frac{1~\mathrm{GHz}}{\nu_c} 
\end{equation}
where $\nu_c$ is the central frequency of the observations.
For the full bandwidth of \SIrange{4}{12}{GHz}, $\nu_c =$ \SI{8}{GHz} and $\mathrm{FWHM}_\mathrm{PB} =$ \ang{;5.25;}, which makes the field of view, FOV = \ang{;21;}.
In addition, we selected a cell size of \ang{;;0.7} and scales sizes of [0, 5, 10, 15, 20, 25, 50, 100, 200] pixels, with a small scale bias 0.9.
The scale size 0 represents a point source, whereas the larger sizes represent wider features that are dominant in the image.
The small scale bias is ``a numerical control to bias the solution towards smaller scales''\footnote{\url{https://casa.nrao.edu/casadocs/casa-5.5.0/global-task-list/task_tclean/about}}.
Scales that were too large were ignored by the software.

The MS-MFS algorithm can also calculate the spectral index of the intensity map by modelling the spectrum of each flux component. 
After obtaining the spectral index maps, we manually corrected them for the primary beam response with the task \emph{widebandpbcor} (which is not yet implemented in \emph{tclean} for wide-band imaging).
In order to do that, we fed the task with the raw data as well as the basename of our imagefiles and selected the action 
\emph{calcalpha}, 
which recalculates and corrects for PB response only the spectral index map and its error\footnote{\url{https://casa.nrao.edu/casadocs/casa-5.5.0/global-task-list/task_widebandpbcor/about}}.
Next, we applied a mask of 5$\sigma$ (which $\sigma$ for BSBD43 is \SI{0.06}{mJy.beam^{-1}} and \SI{0.2}{mJy.beam^{-1}} for BSBD60) in order to bring forth the most significant areas of emission (Fig.~\ref{fig:bd_spindx_all} top).
We repeated the steps for the spectral index error maps (Fig.~\ref{fig:bd_spindx_all} bottom).

As it can be seen from Fig.~\ref{fig:bd_spindx_all}, both shocks seem to emit a mixture of TE and NTE.
In particular, for BSBD43, a spectral index of $\sim-$1.5 appears to be consistent across the shock.
Most of the areas of the shock that have very negative values ($<-$2) are also the areas where the error is $>$1.
Similarly, the spectral index of BSBD60 throughout the entire shock fluctuates around $-$1, while the errors are very small around the main area of emission.

\subsubsection{Single-Dish Data}
%%%%%%%%%%%%%%%%%%%%%%%%%%%%%%%%%%%%%%%%%%%%%%%%%%%%%%%%%%%%%%%%%%%%%%%%%%%%%%%%
Effelsberg single-dish data has significantly lower resolution than VLA but is sensitive to emission at all spatial scales.
The intensity maps from the Effelsberg observations of BSBD43 and BSBD60 are presented in Fig.~\ref{fig:eff_int}, showing larger-scale features of the emitting regions than what is accessible with the VLA. 
The emission is partially resolved for BSBD43, revealing an extended arc of emission along the bow shock.
For BSBD60, the Bubble is about the same size as the Effelsberg beam at \SI{4.5}{GHz}, and so no details are visible.
Here again the arc-shaped emission from the suspected bow shock is detected with Effelsberg, and traces the optical emission quite well \citepads{2019A&A...625A...4G}.
In both observations, the H~\textsc{ii} regions or bright-rimmed clouds outside of the bow shock are not resolved and contribute to the bow shock's emission.

%%%%%%%%%%%%%%%%%%%%%%%%%%%%%%%%%%%%%%%%
%%%%%%%%%%%%%%%%%%%%%%%%%%%%%%%%%%%%%%%%
\subsection{Morphology}\label{subs:morph}
%%%%%%%%%%%%%%%%%%%%%%%%%%%%%%%%%%%%%%%%
%%%%%%%%%%%%%%%%%%%%%%%%%%%%%%%%%%%%%%%%
\begin{table*}
\caption{Fluxes within a 5$\sigma$ contour of Fig. \ref{fig:all_int}, for the wide-band VLA data, the smoothed C-band, Effelsberg and feathered data. 
}\label{tab:five_sig}
%\centering
\vspace{-0.25cm}
%%%%%%%%%%%%%%%%%%%%%%%%%%%%%%%%%%%%%%
\begin{threeparttable}
\begin{tabular}{crrrrrrrr}
\toprule
\midrule
%%%%%%%%%%%%%%%%%%%%%%%%%%%%%%%%%%%%%%
$\nu$  
& \multicolumn{4}{c}{$F_\nu, _\mathrm{BSBD43}$ (mJy)} 
& \multicolumn{4}{c}{$F_\nu, _\mathrm{BSBD60}$ (mJy)}
\\
(GHz) 
& C\&X-band & C-smooth & Effelsberg & feathered 
& C\&X-band & C-smooth & Effelsberg & feathered
\\
%%%%%%%%%%%%%%%%%%%%%%%%%%%%%%%%%%%%%%
\midrule
%%%%%%%%%
4.5  
%%%% 
& 126.9 
& 126.4
& 214.5 
& 377.1
%%%%
& 716.6
& 662.8
& 148.7 
& 587.4 \\
%%%%%%%%%
%%%%%%%%%
4.7 
%%%%
& n/a 
& {120.3}   
& 219.7 
& {381.9}  
%%%%
& n/a 
& {614.4}  
& 151.0  
& {561.6}\\
%%%%%%%%%
%%%%%%%%%
5.0 
%%%%
& n/a 
& {110.6}   
& 215.9  
& {336.8}  
%%%%
& n/a 
& {614.7}  
& 149.7   
& {561.4} \\
%%%%%%%%%
%%%%%%%%%
5.4 &
%%%%
n/a &
{101.9}   &
217.9  &
{328.1} &
%%%%
n/a &
{575.6}  &
150.0  &
{539.4}
\\
%%%%%%%%%
5.5  &  
%%%%
{116.0}  &
\multicolumn{3}{c}{n/a} &
%%%%
{629.7} &
\multicolumn{3}{c}{n/a}
\\
%%%%%%%%%
5.6 &
%%%%
n/a &
{95.0}    &
220.0  & 
{324.7} &
%%%%
n/a &
{542.2}  &
150.0  &
{517.4}
\\
%%%%%%%%%
6.5  & 
%%%%
{86.7} &
\multicolumn{3}{c}{n/a} &
{539.4} &
\multicolumn{3}{c}{n/a}
\\
%%%%%%%%%
6.6 &
%%%%
n/a &
{74.9}   &
217.9 &
{308.5} &
%%%%
n/a &
{517.0} &
150.9  &
{475.1 } 
\\
%%%%%%%%%
7.5  &
%%%%
{59.0}  &
\multicolumn{3}{c}{n/a} &
%%%%
{550.7} &
\multicolumn{3}{c}{n/a}
\\
%%%%%%%%%
7.6 &
%%%%
n/a &
{39.4}    &
218.4 &
{328.7} &
%%%%
n/a &
\multicolumn{3}{c}{n/a}
\\
%%%%%%%%%
8.5  &
%%%%
n/a &
\multicolumn{3}{c}{n/a} &
%%%%
{459.9}  &
\multicolumn{3}{c}{n/a}
\\
%%%%%%%%%
9.5  &
%%%%
{44.5}  &
%%%%
\multicolumn{3}{c}{n/a} &
n/a &
\multicolumn{3}{c}{n/a}
\\
%%%%%%%%%
10.5  &
%%%%
n/a &
\multicolumn{3}{c}{n/a} &
%%%%
{366.2} &
\multicolumn{3}{c}{n/a}
\\
%%%%%%%%%
11.5 &
%%%%
n/a &
\multicolumn{3}{c}{n/a} &
n/a &
\multicolumn{3}{c}{n/a}
\\
%%%%%%%%%
\midrule
$\alpha$   &
{-1.49} $\pm$ 0.08 &
{-2.11} $\pm$ 0.1  &
0.02  $\pm$ 0.11 &
{-0.31 } $\pm$ 0.16 &
%%%%%%
{-0.76} $\pm$ 0.07 &
{ -0.64 $\pm$ 0.16} &
 0.02 $\pm$ 0.16 &
{-0.54} $\pm$ 0.16
\\
%%%%%%%%%%%%%%%%%%%%%%%%%%%%%%%%%%%%%%
%%%%%%%%%%%%%%%%%%%%%%%%%%%%%%%%%%%%%%
\bottomrule
\end{tabular}
\begin{tablenotes}
      \small
      \item {\textbf{Notes.} 
        The uncertainties are assumed to be 5\% of the flux (calibration error, \citeads{2017ApJS..230....7P}).
        Note that for the Effelsberg data on BSBD60, the VLA 5$\sigma$ contour is smaller than the beam size in some directions and so some of the total flux is not counted in this table. 
        It was also not possible to separate emission from the H~\textsc{ii} region to the NE and the bow shock itself.}
    \end{tablenotes}
  \end{threeparttable}
\end{table*}

BSBD43 (Fig.~\ref{fig:all_int}, left) appears to have a clear bow-like form, with main emission areas at the apex and western part of the shock.
At the northern side of the shock there is a much brighter source.
The data shows that it could be a photoionised region. 
Previous study (\citeadsalias{2021MNRAS.503.2514B}) reports the source as the H~\textsc{ii} region 82.454+02.369 \citepads{1989ApJS...71..469L}, although the spatial resolution of the \citeauthorads{1989ApJS...71..469L} observations does not allow to distinguish between emission from this bright source and the bow shock.

The star itself is located south of the shock and we have not detected it, so it is represented instead with an asterisk in Fig.~\ref{fig:all_int}.
Further south, two more radio point sources can be seen, NVSS~J203418+435650  and WSRTGP~2031+4343 \citepads{1996ApJS..107..239T}.
In the same figure, we have overplotted contours of the brightest emission from the catalogues NVSS at \SI{1.4}{GHz}
\citepads[cyan,][]{1998AJ....115.1693C} and WISE \citepads[blue,][]{2010AJ....140.1868W} at \SI{22}{\um}.
From this, we can see that a significant part of the eastern extension of the bow shock detected by WISE and NVSS is not detected by our VLA observations.
This part is also visible in our single-dish observation (Fig.~\ref{fig:eff_int}, left), but in that case every other detail is not present.

For BSBD60, the details seen in our high-resolution radio image (Fig.~\ref{fig:all_int}, right) correspond very well with the narrow-band nebular emission maps seen in optical with e.g. HST \citepads{2002AJ....124.3313M}\footnote{see also: \url{https://hubblesite.org/contents/media/images/2016/13/3725-Image.html}}.
The bright arc on the North side of the Bubble corresponds to the brightest rim of the optical counterpart, and the bright-rimmed pillar seen projected onto the Bubble is the brightest emission region in radio.
Further bright-rimmed clouds photoionised by \bdsixty to the NNE of the Bubble are also detected with the VLA.
Again, the star itself, \bdsixty, was not detected and is represented instead with an asterisk in Fig.~\ref{fig:all_int}.
NVSS and WISE contours are also overplotted.
We can see that their brightest emission matches with our observation.
In contrast, the details of the bow shock are not visible in our Effelsberg observations (Fig.~\ref{fig:eff_int}, right), where only the outer part of the Bubble is visible.

Both images were centred at the apex of each individual source, however the BSBD60 emitting region does not extend as far from the phase centre as BSBD43 ($\sim$\ang{;;160} as opposed to $\sim$\ang{;;300} total width of BSBD43), contributing to better resolution of our observations. 
This is primarily due to the observing limits of the VLA.
In particular, for our chosen D configuration, for C-band those limits are \ang{;;12}--\ang{;;240}, while for X-band are \ang{;;7.2}--\ang{;;145}.
For sources with emission structures on scales larger than those limits the VLA cannot detect any emission, while for smaller structures the array acts like a single-dish instrument, smoothing the image to a lower resolution\footnote{\url{https://science.nrao.edu/facilities/vla/docs/manuals/oss/performance/resolution}}. 
In our case, BSBD60 is within the observational limits of the array, but BSBD43 is not, resulting in loss of emission and resolution.

As a final remark on the morphology of the sources, we determine the standoff distance of the bow shocks, which can be calculated from \citepads{1970DoSSR.194...41B}:
%%%%
\begin{equation} \label{eq:r_so}
    R_{\mathrm{SO}} = \sqrt{\frac{\dot{M}_w \nu_\infty}{4\pi\rho_{\mathrm{ISM}}(\nu_\star^2+c_s^2)}}
\end{equation}
%%%%
where $\dot{M}_w$ is the mass-loss rate of the star, $\rho_{\mathrm{ISM}}$ the density of surrounding \gls{ism}, $\nu_\star$ the star velocity and $c_s$ speed of sound in the surrounding \gls{ism}.
This characteristic scale of the bow shock gives the distance from the star where the ram pressure of the stellar wind equals the ram and thermal pressure of the \gls{ism} the star travels through.
Given that $n_\mathrm{H}$ is $\approx$\SI{15}{cm^{-3}} for BSBD43 and $\approx$ \SI{50}{cm^{-3}} for BSBD60, assuming standard \gls{ism} abundances, the $\rho_\mathrm{ISM}$ can be calculated from \citepads{2019A&A...625A...4G}:
%%%
\begin{equation} \label{eq:n_h}
n_\mathrm{H} = 0.71\, \rho_\mathrm{ISM} / m_\mathrm{p}
\end{equation}
%%%
where m$_\mathrm{p}$ the proton mass.
Combined with Eq. \ref{eq:r_so}, R$_\mathrm{SO}$ is \ang{;2.72;} for BSBD43 and \ang{;0.42;}  for BSBD60.
This can be compared to the angular distance of the bow shock from the star, as measured from the data image in CASA. 
This results in $\sim$\ang{;4;}, for BSBD43 and $\sim$\ang{;0.6;}, for BSBD60.
In both cases, the calculated R$_\mathrm{SO}$ is about 0.7 of the measured value.
This is in agreement with \citetads{2020MNRAS.493.4172S}, who find that R$_\mathrm{SO}$ is consistent with the location of the wind termination shock, not the swept up \gls{ism}. 

%%%%%%%%%%%%%%%%%%%%%%%%%%%%%%%%%%%%%%%%
%%%%%%%%%%%%%%%%%%%%%%%%%%%%%%%%%%%%%%%%
\subsection{Spectral Analysis}\label{subs:freq_an}
%%%%%%%%%%%%%%%%%%%%%%%%%%%%%%%%%%%%%%%%
%%%%%%%%%%%%%%%%%%%%%%%%%%%%%%%%%%%%%%%%

\begin{table}
\caption{
Flux as a function of photon energy/frequency for BSBD43 and BSBD60, including radio data from this work and the literature, and X-ray and $\gamma$-ray upper limits from the literature.
}
\label{tab:flux_comb}
\begin{threeparttable}
\centering
\vspace{-0.25cm}
\begin{tabular}{llrrr}
\toprule
\midrule
Energy 
& Frequency 
& \multicolumn{2}{c}{$\nu F_{\nu}$} 
& Refs. \\
(eV) &
(GHz) &
\multicolumn{2}{c}{($10^{-12}$ \si{erg.cm^{-2}.s^{-1}}) }&
\\
&
&
BSBD43 &
BSBD60\\
\midrule
%%%%%%%%%%%
\num{5.8e-6} &
1.40  & 
0.0014 & 
n/a & 
(1)\\
%%%%%%%%%%%
\num{8.3e-6} &
2.00    & 
0.014 & 
n/a & 
(2)\\
%%%%%%%%%%%
\num{1.9e-5} &
4.50   &  
0.007 & 
0.037 &
(3)\\
%%%%%%%%%%%
\num{2.3e-5} &
5.50   &  
0.0078 &
0.0385 &
(3)\\
%%%%%%%%%%%
\num{2.7e-5} &
6.50   & 
0.0078 &
0.0432 &
(3)\\
%%%%%%%%%%%
\num{3.1e-5} &
7.50   &
0.005175 &
0.04485 &
(3)\\
%%%%%%%%%%%
\num{3.5e-5} &
8.50   &
n/a  &
0.04335 &
(3)\\
%%%%%%%%%%%
\num{3.9e-5} &
9.50   &
0.00437 &
n/a &
(3)\\
%%%%%%%%%%%
\num{4.3e-5} &
10.5   &
n/a &
0.04515 &
(3)\\
%%%%%%%%%%%
\num{1.7e3}     
& \num{4.11e8}        
& n/a 
& 0.01 
& (4),$3\sigma$* \\
%%%%%%%%%%%
\num{2.2e3} 
& \num{5.30e8}  
& 0.015 
& n/a 
& (5),$2\sigma$*\\
%%%%%%%%%%%
\num{4.2e8} 
& \num{e14}
& 1.6022 
& n/a       
& (6),$2\sigma$*	\\
%%%%%%%%%%%
\num{3.1e9} 
& \num{7.40e14} 
& 0.5287 
& n/a       
& (6),$2\sigma$*\\
%%%%%%%%%%%
\num{2.3e10} 
& \num{5.60e15} 
& 1.6823 
& n/a       
& (6),$2\sigma$* \\
%%%%%%%%%%%
\num{1.7e11} 
& \num{4.10e16} 
& 1.9066 
& n/a       
& (6),$2\sigma$* \\
%%%%%%%%%%%
\bottomrule
\end{tabular}

\begin{tablenotes}
      \small
      \item \textbf{Notes.} *:upper limits
      \item \textbf{References.} 
      (1) \citetads{2010A&A...517L..10B}, 
      (2) \citetads{2021MNRAS.503.2514B}, 
      (3) this work,
      (4) \citetads{2020MNRAS.495.3041T}, 
      (5) \citetads{2016ApJ...821...79T}, 
      (6) \citetads{2014A&A...565A..95S}
      
    \end{tablenotes}
  \end{threeparttable}    
\end{table}

\begin{figure*}
    \centering
    \includegraphics[width=\columnwidth, trim={0 0.5cm 0 0.5cm}, clip]{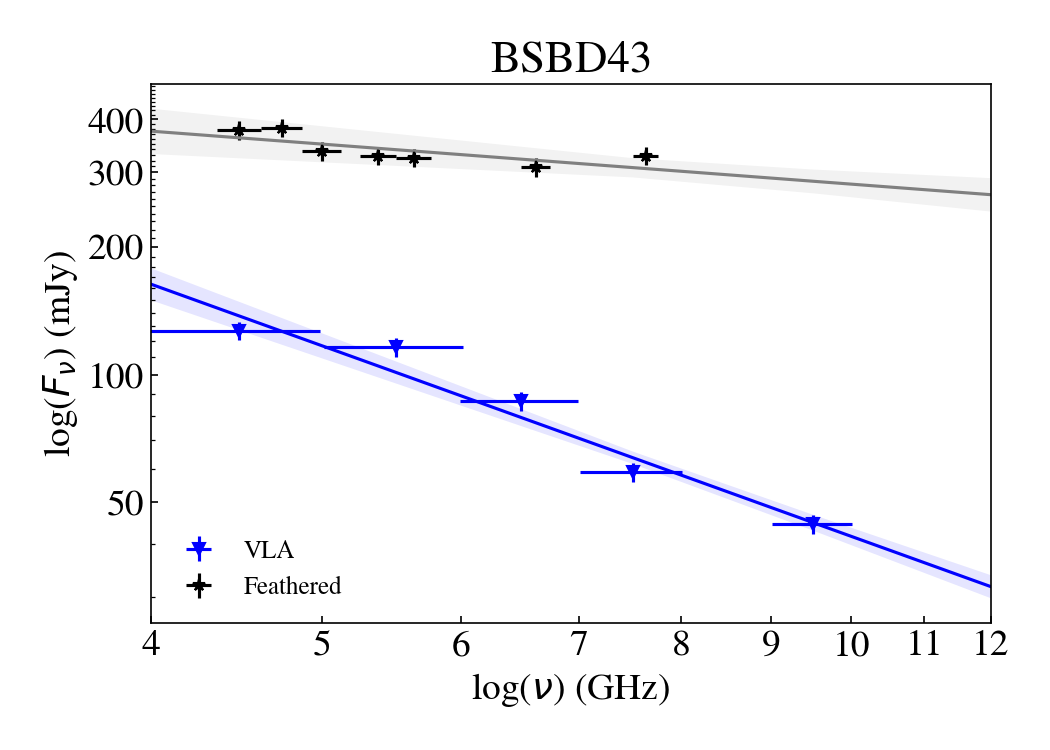}\hfill
     \includegraphics[width=\columnwidth, trim={0 0.5cm 0 0.5cm}, clip]{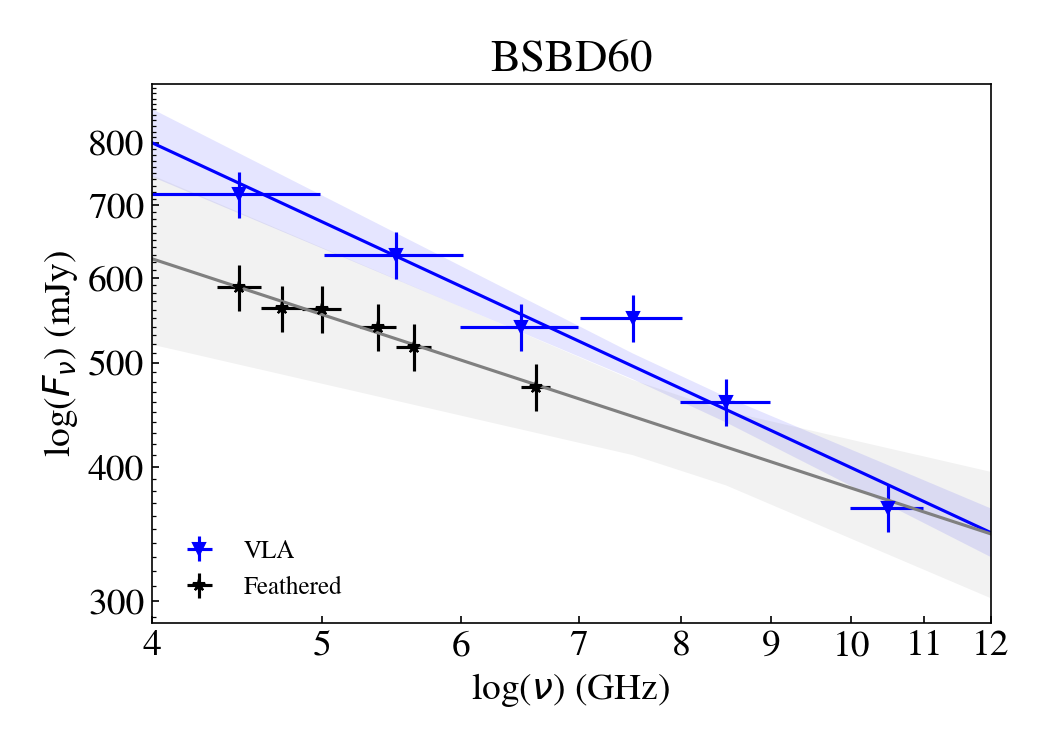}
    
    \caption{
      Power-law fits to the flux within the 5$\sigma$ contour of Fig.~\ref{fig:all_int} from the VLA C\&X-band data (blue) and for the feathered VLA+Effelsberg dataset (black) in Table~\ref{tab:five_sig}.
      The horizontal error-bars show the width of each sub-band, and the vertical error-bars show the flux uncertainty (1$\sigma$).
      The shaded regions show the 1$\sigma$ bounds of the fit.
      }
    \label{fig:sed}
\end{figure*}

\subsubsection{Interferometric Data}\label{subsubsec:ifrd}
The spectral index appears to be fluctuating throughout the entire shock for both targets. 
In order to investigate this fluctuation more quantitatively, we split the data into smaller sub-bands of \SI{1}{GHz} width and calculated their spectral energy distribution (SED) within each bow shock.

A single VLA sub-band has a width of \SI{128}{MHz}.
Therefore, by combining 8 sub-bands and following similar steps for imaging the full band-width, we can obtain an image of $\sim$\SI{1}{GHz}-width.
First, we chose the shortest common $uv$-range (16k$\lambda$), then corrected for the primary beam and smoothed to the largest common beam, so we can compare them equally.
Following, with the help of the CASA task, \emph{imstat}, we integrated all of the flux within the 5$\sigma$ contours, as shown in Fig.~\ref{fig:all_int}, for each bow shock (excluding other sources) as a function of frequency, and calculated the spectral index of the diffuse emission.
The fluxes are presented in Table~\ref{tab:five_sig}.

Both in the case of TE and NTE, the SED follows a power-law:
\begin{equation} \label{eq:pow_law}
F_{\nu} \propto \nu^{\alpha} 
\end{equation}
with $\alpha$ being the spectral index, its values ranging from $-0.1$ to +0.2 for TE, while for NTE it can reach negative values down to $-0.8$.
We used the function \emph{curve\_fit} from the SciPy python library in order to fit the fluxes to the power-law of Eq. \ref{eq:pow_law} and calculate the average spectral index, assuming that all emission inside the limiting contour comes exclusively from the bow shock.
The calculation yielded:
$\alpha_{\mathrm{BSBD43}} = -1.49 \pm 0.08$ and $\alpha_{\mathrm{BSBD60}} = -0.76 \pm 0.07$ (Table~\ref{tab:five_sig}), although note that the obtained spectral index for BSBD43 is sensitive to the chosen dataset (see below for feathered data).

The fit to the VLA C\&X-band data is plotted at Fig.~\ref{fig:sed}, where the errors of the flux are 5\% calibration uncertainty \citepads{2017ApJS..230....7P} and the frequency range is $\pm$\SI{0.5}{GHz}. 
The upper and lower limit envelopes of the fit are determined from:
\begin{align}
 &\nu^{\alpha}F_0-\mathrm{min}[\nu^{\alpha-\alpha_\mathrm{err}},\nu^{\alpha+\alpha_\mathrm{err}}](F_0+F_\mathrm{err}), \\
 &\nu^{\alpha}F_0+\mathrm{min}[\nu^{\alpha+\alpha_\mathrm{err}},\nu^{\alpha-\alpha_\mathrm{err}}](F_0-F_\mathrm{err})     \;,
\end{align}
where $\alpha$ and $\alpha_\mathrm{err}$ is the resulting spectral index from the fitting and its uncertainty, while $F_0$ and $F_\mathrm{err}$ are the flux normalisation and its uncertainty.

\subsubsection{Single Dish}
The data was split into sub-bands of $\sim$\SI{200}{MHz} width to match the centre frequencies of the single VLA sub-bands.
Their typical 1$\sigma$ noise levels were about 6-\SI{6.5}{mJy}.
From those frequency ranges, the ones where the final images in both cases (single-dish and interferometric data) had acceptable noise-to-signal ratios were chosen, namely 4.5, 4.7, 5, 5.4, 5.6, 6.6 and \SI{7.6}{GHz}. 

\begin{figure}
    \centering
    \includegraphics[width=0.9\columnwidth]{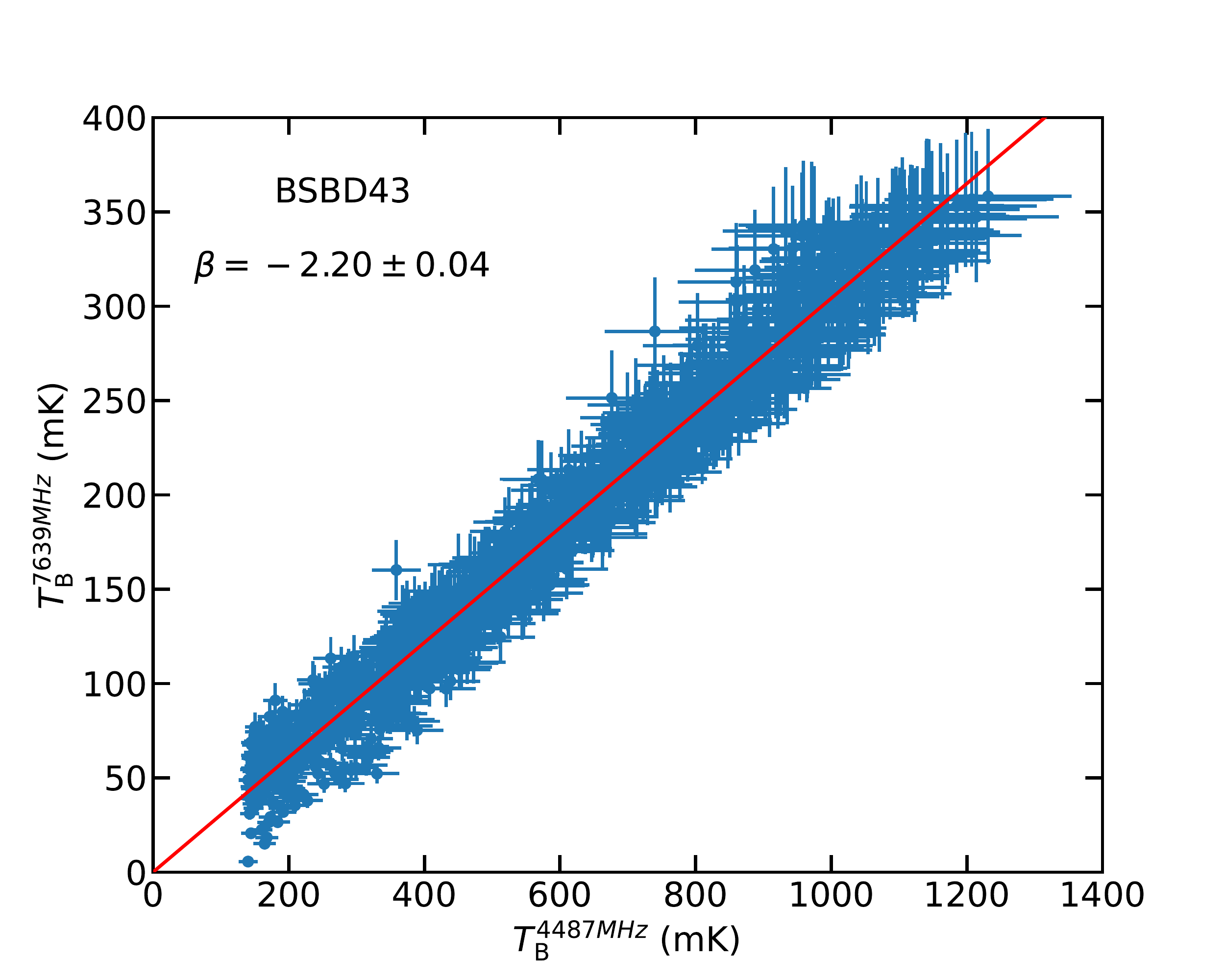}
    \includegraphics[width=0.9\columnwidth]{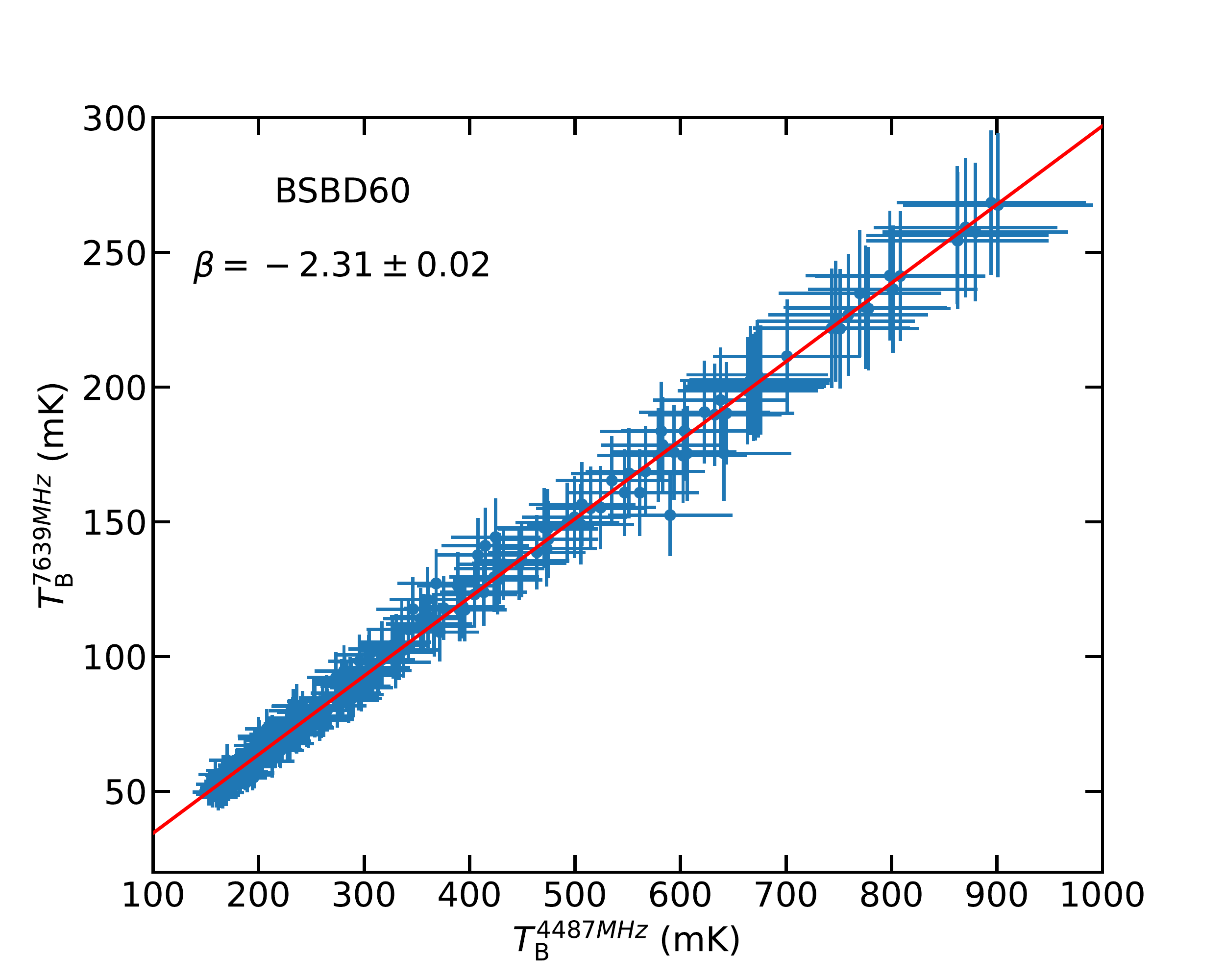}
    \caption{TT plot of the bow shock of BSBD43 (top) and BSBD60 (bottom) using the Effesberg continuum images at 4.487 GHz and 7.639 GHz. 
    }
    \label{fig:tt}
\end{figure}

In order to derive the spectral index from the single-dish observations, we smooth all the images to the common angular resolution of \ang{;;152}. The temperature-temperature (TT) plot is widely used in single-dish observations to estimate the spectral index of extended sources  \citepads{1962MNRAS.124..297T}, and is thus adopted as our approach. 
The TT plot between the continuum emission at \SI{4.487}{GHz} and \SI{7.639}{GHz} is shown in the top panel of Fig.~\ref{fig:tt} where we only take the peak flux density of $>$\SI{50}{mJy} at \SI{4.487}{GHz} into account and a 10\% flux calibration has been included. Defining $T_{\rm B}(\nu)\sim \nu^{\beta}$ where $T_{\rm B}(\nu)$ is the brightness temperature at frequency $\nu$ and $\beta$ is the brightness temperature spectral index, we obtain $\beta=-2.20\pm0.04$ and $\beta=-2.31\pm0.02$ for BSBD43 and BSBD60 by the linear fit to the data points, respectively. 
We also apply this method to different pairs of the Effelsberg continuum images, which vary within in the range from $-2.30$ to $-2.00$ for BSBD43 and from $-2.41$ to $-2.18$ for BSBD60. The variation may arise from the calibration uncertainties.

In order to minimise the zero level uncertainties of the Effelsberg continuum images, we also make use of the background filtering method \citepads{1979A&AS...38..251S} to filter out the emission at the large scale of $>$\ang{;10;}, which could arise from pervasive, diffuse, unrelated emission (i.e., originating in the Galactic disk). 
Then, we perform the linear fit to the data of each pixel to study the distribution of the spectral index. 
Assuming $T_{\rm B}(\nu)=T_{5} \nu^{\beta}$ where $T_{5}$ is the brightness temperature at \SI{5}{GHz}, we use the emcee code \citepads{2013PASP..125..306F} to perform the Monte Carlo Markov chain (MCMC) calculations with the affine-invariant ensemble sampler \citepads{2010CAMCS...5...65G} to fit all the Effelsberg continuum images. The uniform priors are assumed for $T_{5}$ and $\beta$.
The likelihood function is assumed to be e$^{-\chi^{2}/2}$ with 
\begin{equation}
    \chi^{2} = \Sigma_{i} (T_{i, {\rm obs}}- T_{i, {\rm mod}})/\sigma_{i}^{2} \;,
\end{equation}
where $T_{i, {\rm obs}}$ and $T_{i, {\rm mod}}$ are observed and modelled brightness temperatures, and $\sigma_{i}$ is the standard deviation of $T_{i, {\rm obs}}$. The calculations are performed with 20 walkers and 8000 steps after the burn-in phase.

\begin{figure}
    \centering
    \includegraphics[width=0.9\columnwidth]{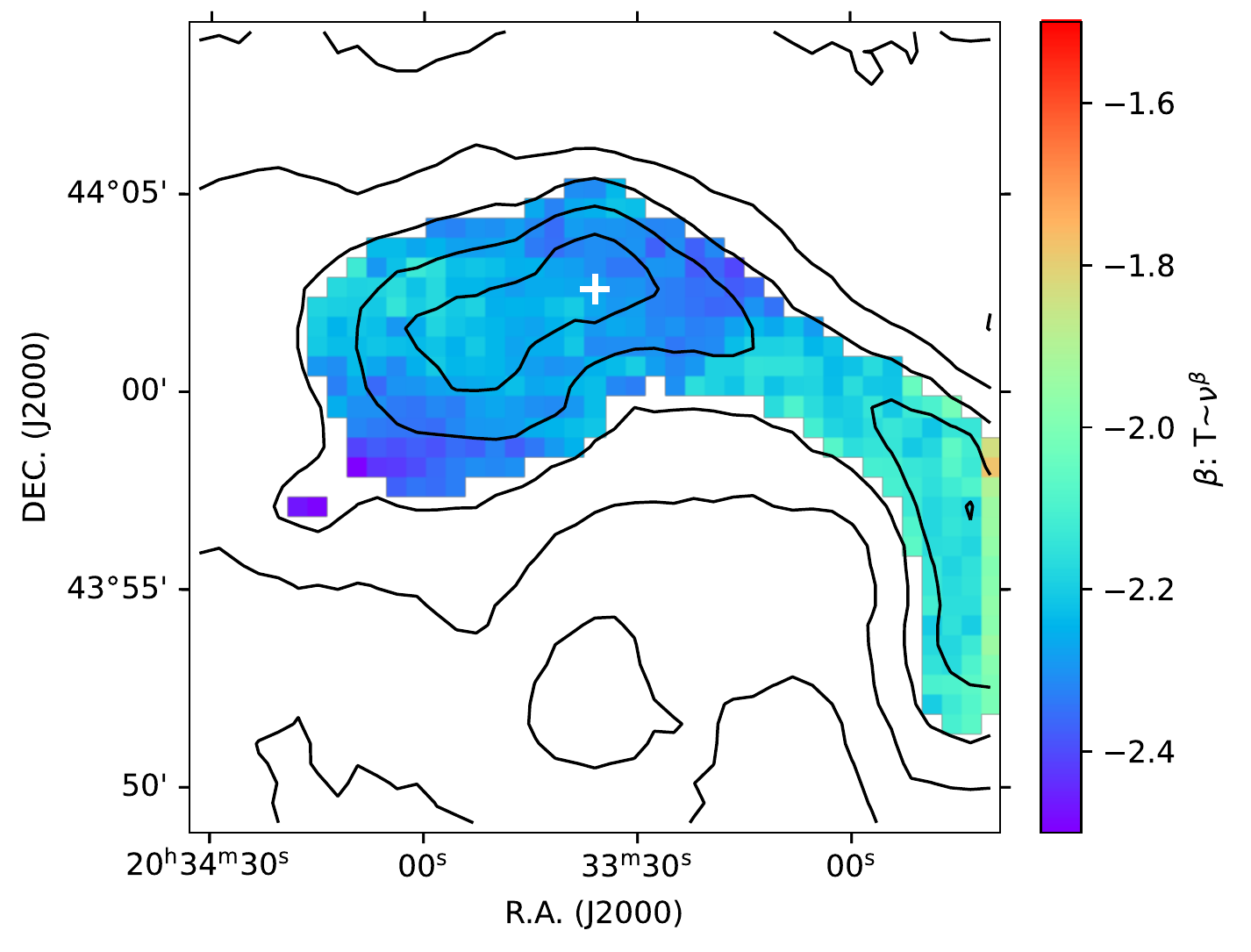}
    \includegraphics[width=0.9\columnwidth]{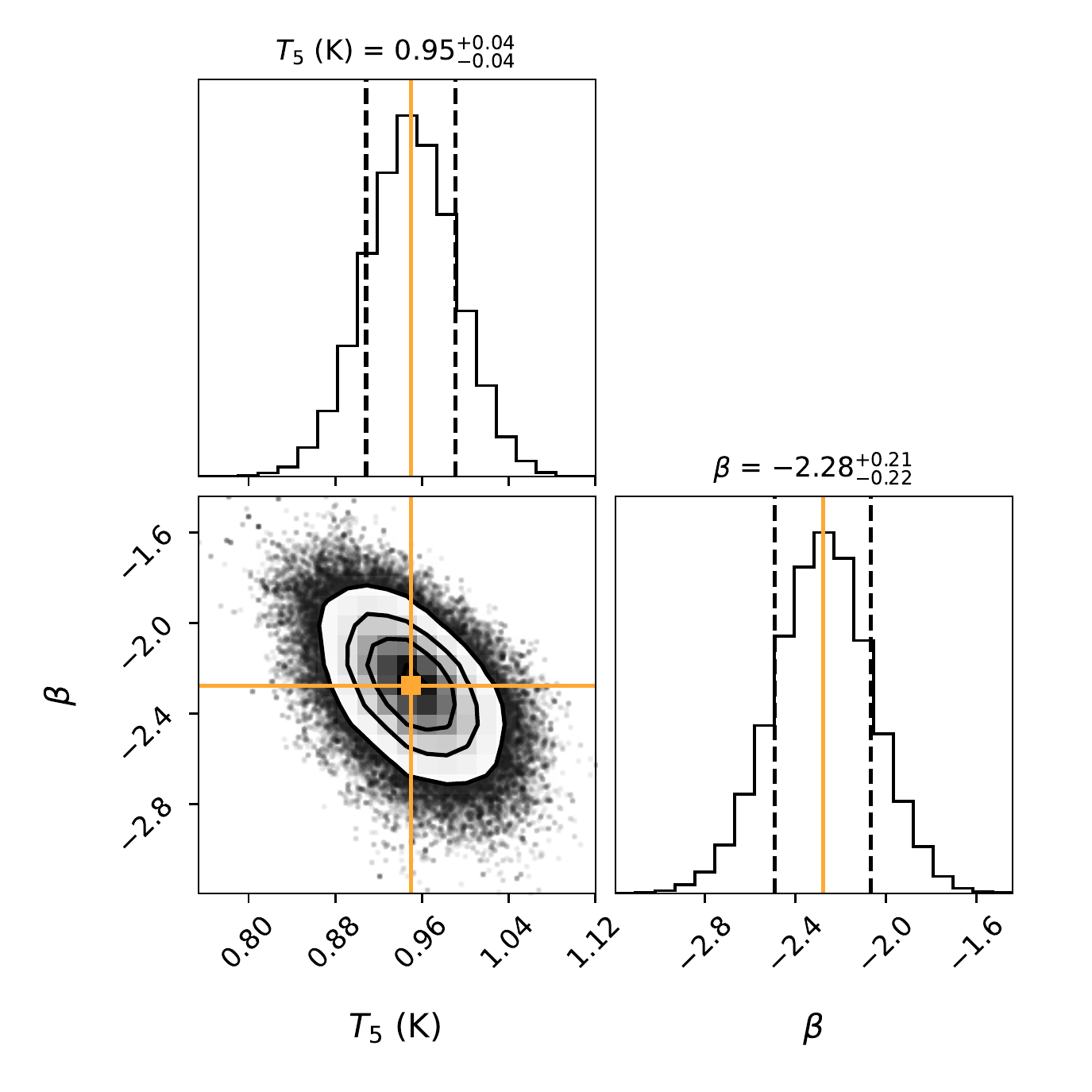}
    \caption{
    \textit{Top}: Spectral index map of the bow shock of BD+43$^\circ$ 3654, as seen with the Effelsberg telescope. As it can be seen, the data are dominated by free-free emission.
    \textit{Bottom}: Posterior probability distributions of the fitted $T_{\rm B}$ and $\beta$ toward the position indicated by the cross in the middle panel.
    The maximum posterior possibility point in the parameter space is shown in orange line. 
    Contours denote the 0.5, 1.0, 1.5, and 2.0$\sigma$ confidence intervals.
    }
    \label{fig:eff_beta}
\end{figure}

\begin{figure}
    \centering
    \includegraphics[width=0.9\columnwidth]{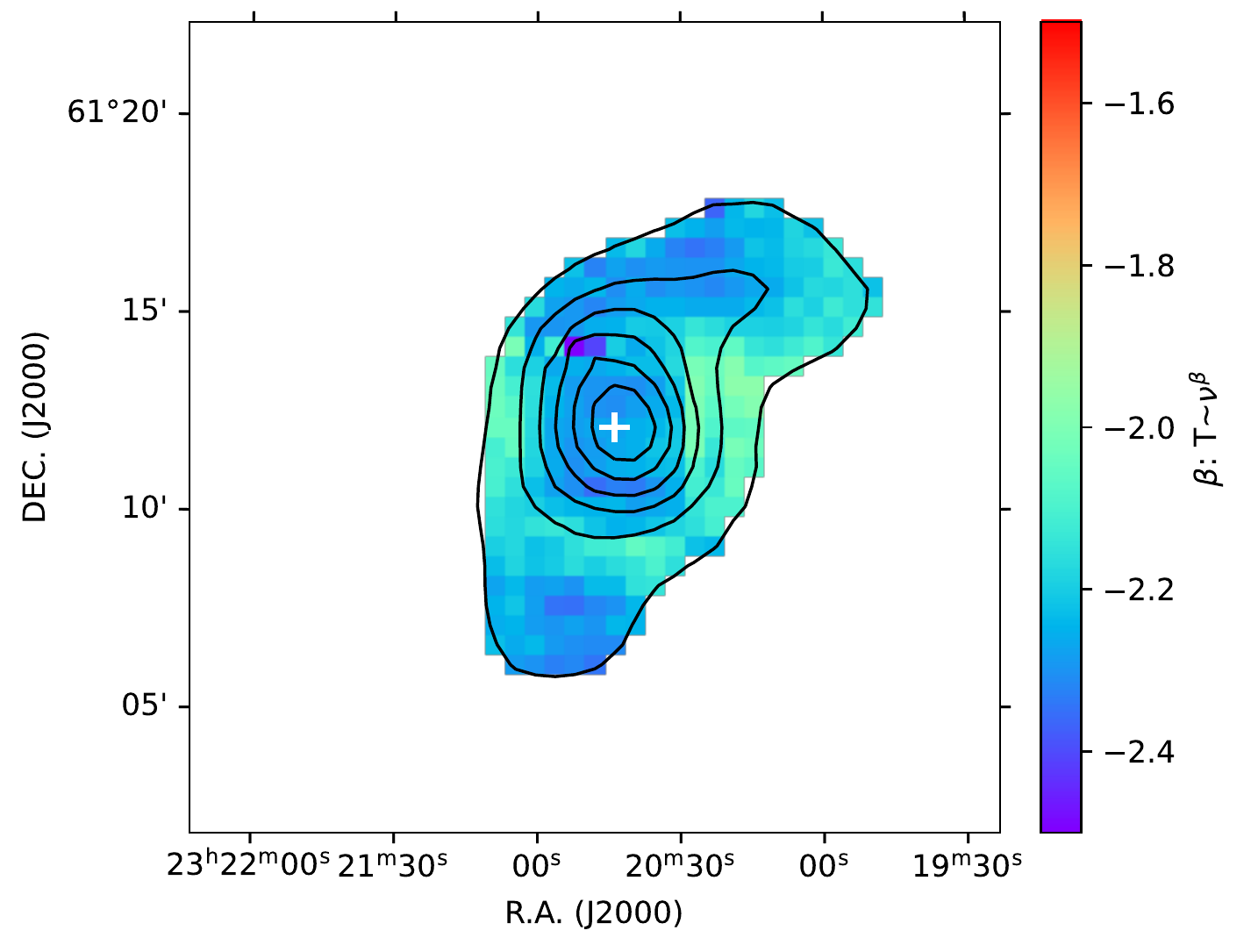}
    \includegraphics[width=0.9\columnwidth]{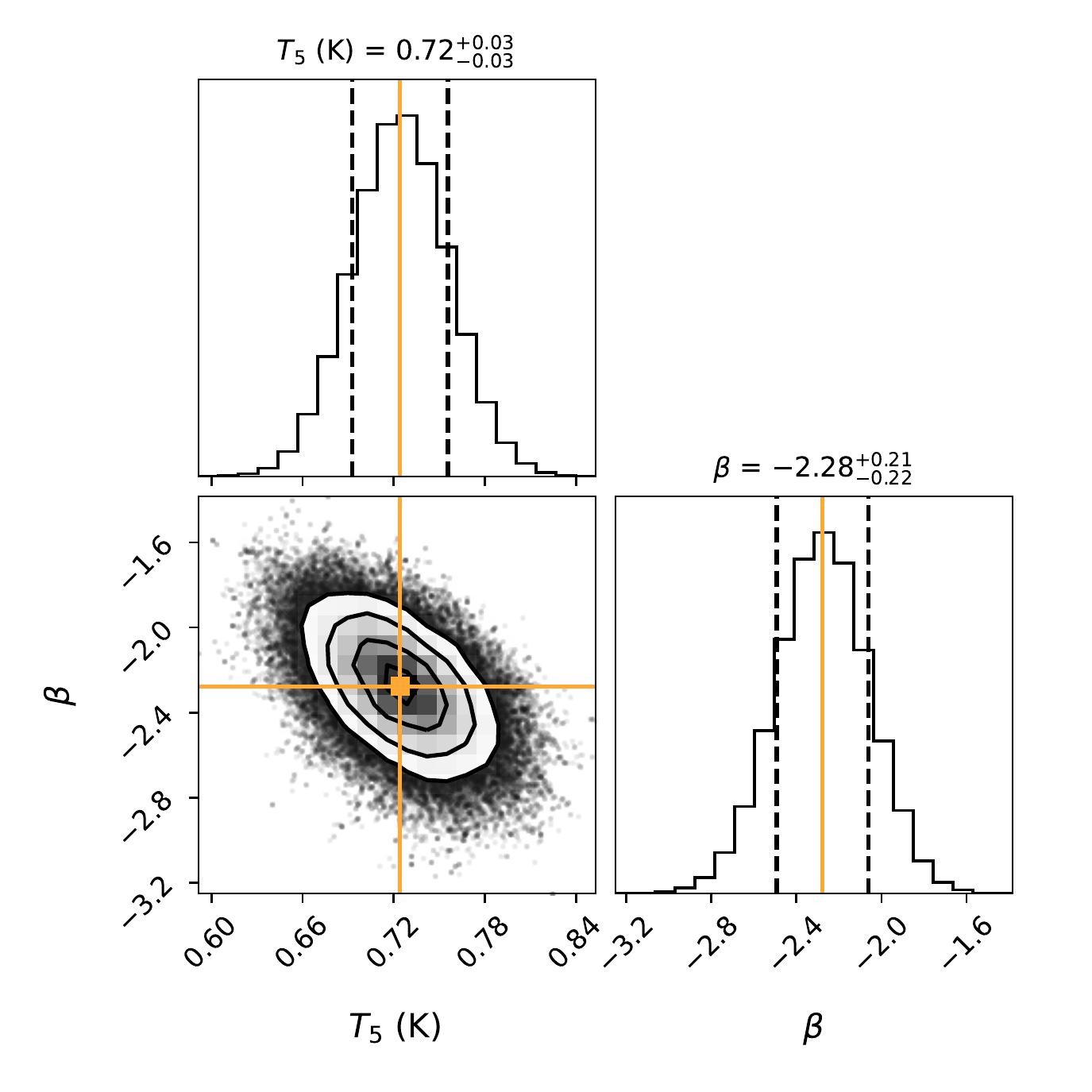}
    \caption{Same as Fig.~\ref{fig:eff_beta} but for BSBD60.
    }
    \label{fig:eff_bsbd60}
\end{figure}

The spectral-index maps are shown in the top panel of Figs.~\ref{fig:eff_beta} and \ref{fig:eff_bsbd60}, and an example of the fitting process is shown in the bottom panel of Figs.~\ref{fig:eff_beta} and \ref{fig:eff_bsbd60}. The spectral index maps of both BSBD43 and BSBD60 suggest $\beta$ of about $-$2.2 similar to the results derived from the TT plot. Taking the typical uncertainties of 0.2 in $\beta$ (see the bottom panel of Fig.~\ref{fig:eff_beta}),
we can infer that all of the emission is consistent with $\beta\approx-2.2$, where $\alpha =\beta+2^($\footnote{The Rayleigh–Jeans law suggests $F_{\nu}\propto \nu^{2}T \propto \nu^{\alpha}$ and the temperature-frequency dependence defines $T\propto \nu^{\beta}$, which gives $\alpha =\beta+2$.}$^)$, corresponding to thermal emission.
This is expected because Effelsberg is sensitive to large-scale diffuse emission from the photoionised H~\textsc{ii} region around the bow-shock-driving star, which is filtered out by the interferometric technique.
Results for BSBD60 are very similar, consistent with thermal emission.

%%%%%%%%%%%%%%%%%%%%%%%%%%%%%%%%%%%%%%%%
%%%%%%%%%%%%%%%%%%%%%%%%%%%%%%%%%%%%%%%%
\subsection{Single-Dish and Interferometry Data Combination}\label{subs:comb}
%%%%%%%%%%%%%%%%%%%%%%%%%%%%%%%%%%%%%%%%
%%%%%%%%%%%%%%%%%%%%%%%%%%%%%%%%%%%%%%%%
In order to combine our VLA data (from now on IFD - Interferometric Data) with our Effelsberg data (from now on SDD - Single-Dish Data), we smoothed the images of each frequency to a common circular beam of FWHM \ang{;;30} for BSBD43, \ang{;;20} for BSBD60 and \ang{;;155} for all SDD.
Next, we combined the IFD and SDD using the CASA task, \emph{feather} \citepads{2019AJ....158....3R}, and converted the results into \si{MJy.sr^{-1}}.
The data for the central frequency \SI{7.6}{GHz} were omitted from the analysis of BSBD60 because the s/n was too low.
All the results are shown in {Figs}.~\ref{fig:comb_all_bd43} and \ref{fig:comb_all_bd60}.

The \emph{feathering} adds to the flux of BSBD43 (Fig.~\ref{fig:comb_all_bd43}), especially to the SE of the apex of the shock, which agrees with the data from WISE and NVSS (Fig.~\ref{fig:all_int} left).
On the other hand, the \emph{feathering} has not added any significant information for BSBD60 (Fig.~\ref{fig:comb_all_bd60}).

Similarly, we calculated the SEDs for the feathered results as well as C-band only for the respective frequencies, and the results are summarised in Table~\ref{tab:five_sig}.
We find that $\alpha$ obtained from feathered maps is much closer to 0 (less negative) than IFD results for BSBD43, and the results are consistent with predominantly TE.
Table~\ref{tab:five_sig} shows that the unphysically negative $\alpha=-1.49\pm0.08$ that we obtained for the total VLA emission from the BSBD43 is changed to $\alpha=-0.31\pm0.16$ with the inclusion of Effelsberg data in the feathered dataset.
Within the uncertainties this is now consistent with mostly TE.
The large difference between the VLA and feathered results indicates that BSBD43 is so diffuse that the VLA is not sensitive to much of the emission, particularly at the higher frequencies (see the respective fluxes in Table~\ref{tab:five_sig}).

For BSBD60, the feathered results are very similar to the IFD.

%%%%%%%%%%%%%%%%%%%%%%%%%%%%%%%%%%%%%%%
% 05 - DISCUSSION
%%%%%%%%%%%%%%%%%%%%%%%%%%%%%%%%%%%%%%%
\section{Discussion}\label{sec:disc}
Wind bubbles are surrounded by H~\textsc{ii} regions \citepads{2015A&A...573A..10M} that are bright thermal radio sources, whereas the outer edge of the wind bubble should be the brightest in NTE \citepads{2018ApJ...864...19D}.
We expect, therefore, the SDD to be sensitive to the more homogeneous and large-scale emission from the surrounding H~\textsc{ii} region, while the IFD to be more sensitive to the NTE, which is less extended and has more small-scale structures.
This is evident in the case of BSBD43, where the addition of the SDD has increased the spectral index, especially the additional areas that seem to appear after the feathering (Fig. \ref{fig:comb_all_bd43}).
This could be due to the fact that SDD is dominated by TE (presumably from the H~\textsc{ii} region) with $S_\nu\propto \nu^{-0.1}$, and that the total flux from Effelsberg is significantly larger than that obtained from the VLA.
This makes the results somewhat difficult to interpret, because the spectral index obtained from the $5\sigma$ contours is consistent with TE in the feathered dataset, whereas from the VLA data alone one would conclude that most of the bow shock is emitting NTE.

There are two possibilities that could explain our results. 
Either the emitting region is too extended for our VLA observations to obtain an accurate spectral index and so the emission is thermal at \SIrange{4}{12}{GHz}, or the VLA is picking out small features in the bow shock that are non-thermal emitters, and these are difficult to see once the bright, large-scale TE from Effelsberg is folded in.
Given the detection of NTE at lower frequencies (\citeadsalias{2021MNRAS.503.2514B}) we favour the latter interpretation.

Contrarily, for BSBD60 (Fig.~\ref{fig:comb_all_bd60}) not much new information is added because, due to the fact that the diameter of the nebula is similar to the Effelsberg beam size, the emitting region is completely unresolved by the SDD.
Our results for BSBD60 are relatively straightforward to interpret: we have detected synchrotron emission from the northern rim of the stellar wind bubble, and a combination of intense TE and NTE from the bright-rimmed cloud projected onto the bubble.

\subsection{Comparison between sources and with literature}
Both BSBD43 and BSBD60 have been studied before in the radio range  \citepads[e.g. \citeadsalias{2010A&A...517L..10B},][]{1982MNRAS.201..429T} but only the latter has not been previously proved to emit NTE in the radio frequencies.
In our analysis, however, it is the only source with self-consistency and this could be mainly due to its spatial extend.
The fact that BSBD60 is smaller in diameter than BSBD43 contributes to a much higher s/n in the VLA images.
This becomes even more evident from a flux comparison of the 5$\sigma$ emission from both bow shocks throughout the bandwidth (Table~\ref{tab:flux_comb}), where BSBD60 is up to 10 times brighter than BSBD43.

On the other hand, our results for BSBD43 do not agree entirely with previous radio studies (\citeadsalias{2010A&A...517L..10B}, \citeadsalias{2021MNRAS.503.2514B}), which suggest that the bow shock emits NTE. 
In our case, the overall spectral index is too negative to be explained by NTE ($-1.6$).
From Fig. \ref{fig:sed}, it can be argued that for frequencies $<$\SI{7}{GHz}, the spectrum is less steep.
In fact, $\alpha_{4-7GHz}=-0.83 \pm 0.19$, which agrees with the results of \citeadsalias{2021MNRAS.503.2514B} for contours of $>$\SI{2.3}{mJy.beam^{-1}}.
This could be due to loss of flux in higher frequencies from the short-spacing problem.

\subsection{Non-detection of the stars}
An additional comment can be made regarding the non-detection of the stars.
\citetads{1975MNRAS.170...41W} states that for a star with a mass-loss $\dot{M}$, at radio and infrared frequencies $\nu$, the total observed flux can be calculated from the following:
\begin{equation}\label{eq:s_nu}
S_{\nu} = 23.2
\left(
\frac{\dot{M}}{\mu \upsilon_\infty}
\right)^{4/3} 
\frac{\nu^{2/3}}{d^2}
\gamma^{2/3} g^{2/3}
Z^{4/3}  \, \mathrm{Jy}\;.
\end{equation}
Here, for an ionised gas with \gls{ism} abundances, the ratio of electrons to ions is $\gamma\approx1.2$, the Gaunt factor is, with 20\% accuracy, $g\approx1.2$ \citepads{1986rpa..book.....R}, the mean atomic gas weight is $\mu=0.61$ and $Z=1$.
The distance $d$ is measured in kpc, wind speed $\upsilon_\infty$ is in \si{km.s^{-1}}, and $\dot{M}$ is in \si{M_\odot.yr^{-1}}.
For our observed frequencies, the mean value of the fluxes is calculated to be
$S_\textrm{BSBD43} \approx$ \SI{0.477}{mJy} and
$S_\textrm{BSBD60} \approx$ \SI{0.014}{mJy}.
On the other hand, our observed $3\sigma$ upper limits are \SI{0.18}{mJy} and \SI{0.3}{mJy} respectively.
This implies two different outcomes for our sources.
For BSBD60, our upper limit is much higher than the calculated value, meaning that the observational upper limit is consistent with the expected flux.
For BSBD43, on the other hand, our upper limit implies that we should have detected this source (marginally) at the $2.7\sigma$ level, suggesting that $\dot{M}$ might be overestimated.
A reverse calculation of Eq. \ref{eq:s_nu} using our 3$\sigma$ upper limit results in a mass-loss rate of $\dot{M}\lesssim\SI{4.4e-6}{M_\odot.yr^{-1}}$.

\subsection{Modelling the source of the non-thermal emission}
There are at least two possible sources of NTE in bow shocks, electrons accelerated at the wind termination shock and radiating in the magnetic field of the shock \citepads[e.g.][]{2018ApJ...864...19D}, and/or enhanced synchrotron emission from compression of the background GCR population and magnetic field in the radiative forward shock of the bow shock \citepads{2019A&A...622A..57C}.
Here we explore both possibilities.

\subsubsection{Synchrotron and inverse-Compton emission model}
\begin{figure*}
    \centering
    \includegraphics[width=0.495\textwidth]{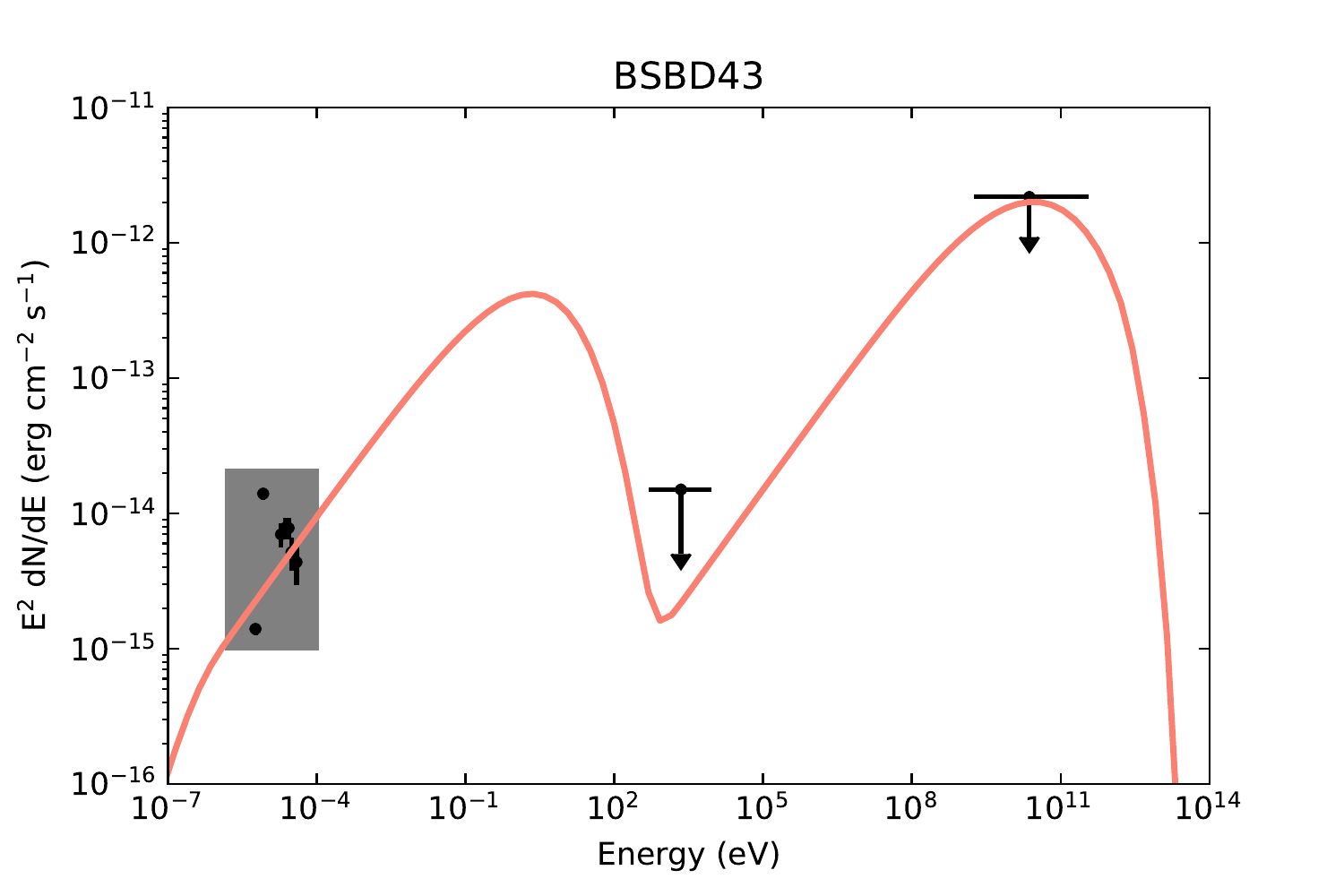}
    \includegraphics[width=0.495\textwidth]{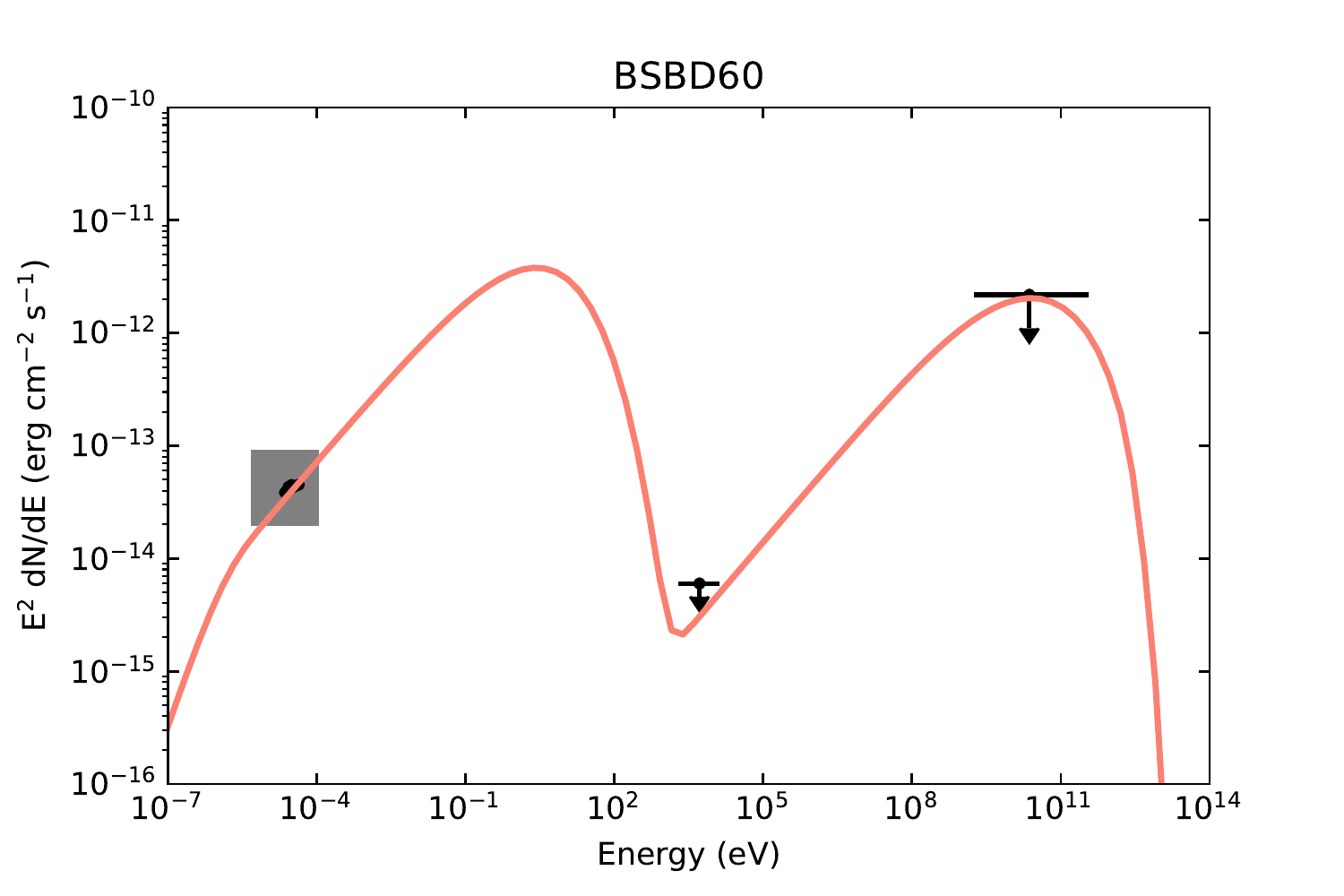}
    \caption{Theoretical modelling of BSBD43 (left) and BSBD60 (right), respectively, by assuming SYN/IC scenario and fitting using the \emph{naima} package.  Fit parameters are in Table~\ref{tab:SYNIC}.}
    \label{fig:modelling_SSC}
\end{figure*}

For both BSBD43 and BSBD60 we only have upper limits for NTE in X-ray and $\gamma$-ray bands, and with only radio detections one cannot predict \textit{a priori} the level of the higher-energy radiation.
Instead, we suppose that the actual X-ray and $\gamma$-ray flux is just below the upper limits quoted in Table~\ref{tab:flux_comb}, use the \textit{naima} package \citepads{2015ICRC...34..922Z} to model the broadband SEDs, and compare the fitting results with the estimated physical conditions within the bow shocks.
Considering the ambiguity of the shape of SED in the radio band, we build a box (grey colour, Fig.~\ref{fig:modelling_SSC}) in the radio band, covering all radio observations in this work. 
We assume a one-zone model with leptonic emission from electrons accelerated in the termination shock of the wind.
The broadband SEDs and results of theoretical modelling are shown in Fig.~\ref{fig:modelling_SSC}.
The X-ray and $\gamma$-ray data are upper limits taken from Table~\ref{tab:flux_comb}, except that for BSBD60 we consider the FERMI-LAT point-source sensitivity as a nominal upper limit because there are no published results.

In the scope of this modelling, it is assumed that the electrons at the wind termination shock have the following energy distribution
\begin{equation}
N(E) \sim  \left(\frac{E}{E_{0}}\right)^{-\Gamma_\mathrm{e}}\exp\left(\frac{E}{E_{c}}\right)\;,\;\; E > E_\mathrm{min} \;,
\label{eq:plecgamma}
\end{equation}
where \textit{E} is the energy of electrons, $N(E)$ the number of electrons per unit energy, $\Gamma_\mathrm{e}$ is the power law index, $E_\mathrm{c}$ and $E_\mathrm{min}$ describe the cut-off and the minimum energy of the electrons, and $E_0$ is a normalisation constant. 
The total energy in relativistic electrons, $W_\mathrm{e}$, is given by
\begin{equation}
W_\mathrm{e} = \int_{E_\mathrm{min}}^\infty N(E)dE \;.
\end{equation}

Considering the possible non-thermal origin of radio emission (\citeadsalias{2010A&A...517L..10B}, \citeadsalias{2021MNRAS.503.2514B}), it can be presumed that the lower energy band, i.e. from radio to X-ray, is due to synchrotron emission from the highly accelerated electrons in the magnetic field, \textit{B}, along the bow shock. 
These electrons also interact with the ambient soft photon fields (from the star and the background) via Inverse Compton (IC) scattering, producing high-energy emission from X-ray to $\gamma$-rays. 
The following radiation fields are considered: 
\begin{enumerate}
\item
Cosmic Microwave Background (CMB; radiation temperature of \SI{2.72}{K}, energy density of \SI{1}{eV.cm^{-3}});
\item
Far-Infrared dust emission (FIR; radiation temperature of \SI{26.5}{K}, energy density of \SI{0.415}{eV.cm^{-3}}); and
\item
Stellar radiation, assuming a blackbody with radiation temperature \SI{40700}{K} and energy density \SI{2.4e-10}{erg.cm^{-3}}  for BSDB43, and radiation temperature \SI{35000}{K} and energy density \SI{9.7e-10}{erg.cm^{-3}} for BSDB60.
\end{enumerate}
The energy density of the stellar field is obtained from the standoff distance of the bow shock and stellar luminosity of each object according to $U_\mathrm{rad}=L/4\pi r^2 c$.
We use $R_\mathrm{SO}\approx$ \ang{;4;} (section 4.2) and $\log L/$L$_\odot= 5.94$ \citepads{2005A&A...436.1049M} for \bdforty, and $R_\mathrm{SO}\approx$ \ang{;0.6;} (section 4.2) and $\log L/$L$_\odot= 5.4$ \citepads{2020MNRAS.495.3041T} for \bdsixty.

The modelling best-fit parameters are presented Table~\ref{tab:SYNIC}. 
In both cases a hard power-law index ($\Gamma_\mathrm{e} \sim 2$) and a cutoff energy ($E_\mathrm{c}\sim$\SIrange{1}{3}{TeV}) are obtained.
The hard spectrum is close to what is expected from DSA processes \citepads[e.g.][]{1983RPPh...46..973D}.
The total evaluated energy of electrons is $W_\mathrm{e}\sim \SI{e47}{erg}$ in both cases.
This can be compared with the total available energy processed through the wind termination shock.
The wind power for each star ($\frac{1}{2}\dot{M} \upsilon_w^2$) determines the total available energy:
\begin{equation}
E_\mathrm{w} = \frac{1}{2}\dot{M} \upsilon_w^2 t_r f_{vol}\;,
\end{equation}
where $t_r$ is the time of residence for an accelerated particle in the shocked wind region, which can be calculated as $t_r \sim R_{SO} / (\frac{1}{4} \upsilon_w $) \citepads{2018A&A...617A..13D}.
Also, $f_\mathrm{vol}$ is the fraction of the spherical region around the star that is responsible for the observable emission.
For \bdforty the radio-bright emission cone has opening angle $\theta\approx$\ang{90} in the upstream direction, and for \bdsixty $\theta\approx$\ang{180}.
If A is the surface area of the cone, then:
$A = 2\pi r^2 (1-\cos\theta/2) \Leftrightarrow f_{vol} = A/(4\pi r^2) = 0.5(1-\cos\theta/2)$, evaluating to $f_{vol} \approx 0.15$ for BSBD43 and $f_\mathrm{vol} \approx 0.5$ for BSBD60.
Using the data from Table~\ref{tab:lit_info}, we find
E$_\mathrm{w, BSBD43}=$ \SI{1.5e47}{erg}
and 
E$_\mathrm{w, BSBD43}=$ \SI{1.8e46}{erg}.
Typically less than $1\%$ of this energy goes to the accelerated electrons for standard shock-acceleration theories, and so we conclude that there is about 100$\times$ (BSBD43) or 1000$\times$ (BSBD60) less energy available than the $W_e$ values obtained by the \emph{naima} models.
This implies that the actual level of non-thermal X-ray and $\gamma$-ray fluxes from these bow shocks is significantly below the upper limits from the literature, with the caveats that our modelling is a simple one-zone calculation with some quite uncertain parameters.

\begin{table}
\caption{The results of the \emph{naima} modelling discussed in subsection 5.2. }
\label{tab:SYNIC}
\begin{threeparttable}

\vspace{-0.25cm}
\centering
\begin{tabular}{rll}
\toprule
\midrule
\textbf{Parameters} & \textbf{BSBD43} & \textbf{BSBD60} \\
\midrule
$\Gamma_{e}$  
& $1.98\pm0.08$              
& $1.96\pm0.06$  
\\
%%%%%%%%%%%%%%%%%%
$E_{c}$\,(TeV) 
& $3.09\pm0.06$              
& $1.65\pm0.07$  
\\
%%%%%%%%%%%%%%%%%%
$B\,$(\si{\mu G})
& $4.61\pm0.41$              
& $19.95\pm0.03$  
\\
%%%%%%%%%%%%%%%%%%
$W_{e}$\,(erg) 
& $(4.7\pm0.2)\times10^{46}$ 
& $(1.2\pm0.1)\times\mathrm{10^{47}}$  \\
\bottomrule
\end{tabular}
\begin{tablenotes}
\item \textbf{Notes.} $\Gamma_e$ is the electron power-law index; 
$E_c$ is the cutoff energy of electrons; 
$B$ is the magnetic field and
$W_e$ is the total energy of the electrons.
\end{tablenotes}
\end{threeparttable}
\end{table}

\subsubsection{Compression of Galactic electrons and protons}

\citetads{2019A&A...622A..57C} attributed the gamma-ray access measured by \textit{Agile} from Orion region to the compression of Galactic CRs in the radiative bow shock of the massive OB-star $\kappa$-Ori moving through a region of a high ambient density. 
Here, we investigate if a similar compression of the Galactic electron and proton distribution can lead to detectable radio and gamma-ray emission.

The number density $n_\mathrm{ad}$ of adiabatically compressed CRs is given by
\begin{equation}
    n_\mathrm{ad}(p) = r^{2/3}n_\mathrm{ISM}(r^{-1/3}p) \mathrm{,}
\end{equation}
where $r$ denotes the total shock-compression ratio and $n_\mathrm{ISM}$ the ambient CR-density at momentum $p$ \citepads{2019MNRAS.482.3843T}. 
We describe the spectrum of hadronic CRs as a power law in total energy, modified at low energies by the particle speed, $\upsilon_p$. 
The electron spectrum is a log-parabola at low energies,
\begin{align}
    N_\mathrm{p}(E) &= N_1\frac{\upsilon_p(E)}{c} (E+mc^2)^{s_1}\label{Eq:BackgroundPRs}\\
    N_\mathrm{e}(E) &= 
        \begin{cases}
            \frac{N_2}{E}\exp\left(-\frac{\log^2({E}/{E_c})}{\sigma}\right)\mathrm{ for }E\leq E_B\\
            N_3E^{s_2} \mathrm{ for }E>E_B
        \end{cases}\label{Eq:BackgroundELs}\mathrm{.}
\end{align}
Both electron and proton background spectra can be directly measured above a few GeV, where solar modulation is unimportant. 
The electron spectrum can be constrained by measurements of diffuse radio emission but the spectral slope of the proton spectrum at low energies remains unclear. 
At high energies the local CR spectrum has an index $s_1 = -2.75$ which gives a spectrum harder than $E^{-2}$ at low energies; we choose the normalisation in accordance with \citetads{2002ApJ...565..280M}. 
For the electron spectrum, we fit the spectra given in \citetads{2011MNRAS.416.1152J} for a galactocentric radius of \SI{6.5}{kpc} with expression (\ref{Eq:BackgroundELs}) as the spectrum shows a continuous change in the index below \SI{4}{GeV}. 
The spectral index at high energies is found to be $s_2 = -3.04$ and $E_B=$ \SI{5}{GeV}. 
These spectra are compatible with direct observations of the electron spectra in the local \gls{ism} by Voyager 1, which also show spectra harder that $s=2.0$ at low energies \citepads{2016ApJ...831...18C}.

It has to be noted that only low-energy particles can be efficiently compressed in the bow shock. 
Higher energy particles have a larger mean-free path $\lambda$ and can not be confined when $\lambda\approx L$, where $L$ is the thickness of the compressed shell. 
Here, we assumed a thickness of $L=0.2R_\mathrm{SO}$. 
For the BSBD43, this means that particles with energies greater $\approx$\SI{15}{GeV} can only inefficiently be trapped.

\begin{figure*}
    \includegraphics[width=0.95\textwidth]{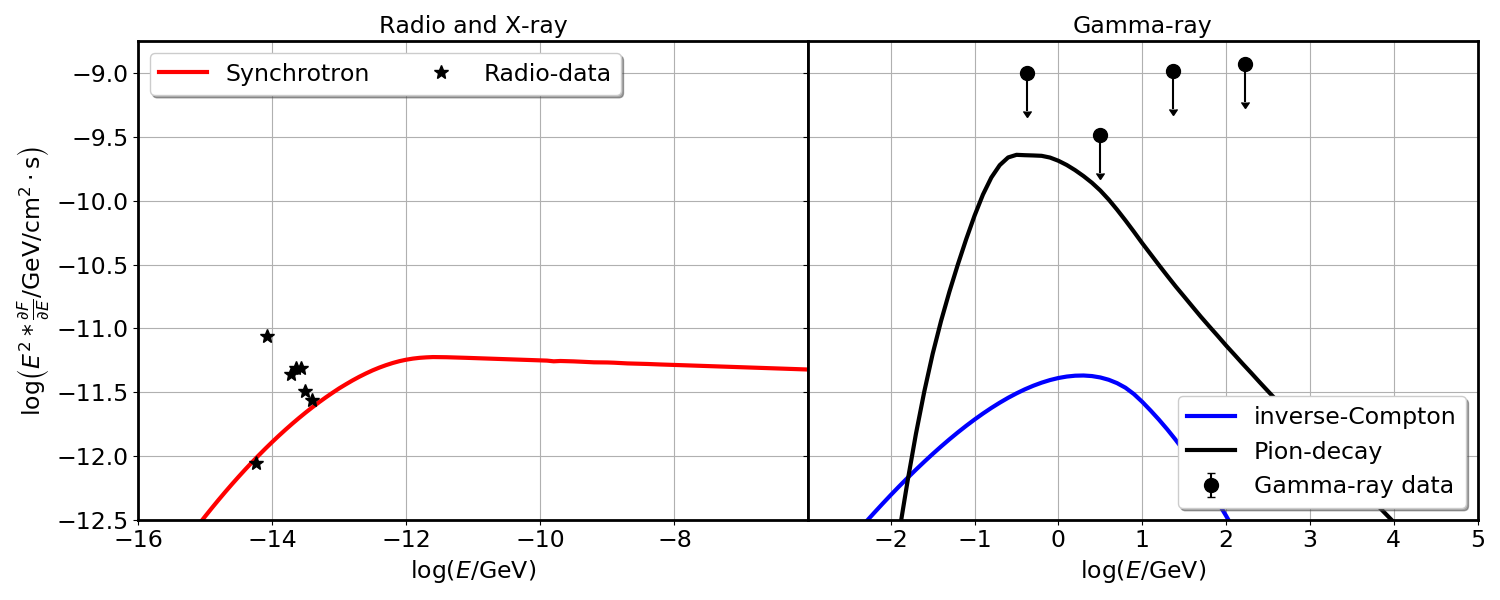}
    \caption{CR compression model for BSBD43, showing the predicted emission for radio (red), inverse-Compton (blue) and pion-decay (black) emission based on the parameters in Table \ref{tab:lit_info}. The measured radio flux is denoted with stars and the gamma-ray upper limits by round markers.}
    \label{fig:BD43_compression}
\end{figure*}

The bow shock itself is likely radiative.
Depending on the diffusion coefficient, even low-energetic CRs might experience a compression ratio larger than the canonical value of 4, which is expected for a strong shock. 
Given the total compression due to radiative cooling of the downstream gas, values of $\Gamma_e\approx1.25$ might not be unreasonable. 
If the termination shock of the inner wind is the place of particle acceleration, local effects might significantly soften the electron spectra to values of $\Gamma_e\gg2$ \citepads{1985ApJ...298..684W}.

In addition to the parameters given in Table~\ref{tab:lit_info}, we assumed a compression ratio at the forward shock of $r=20$ and an ambient magnetic field strength of $B_0=$ \SI{5}{\mu G}. 
Since not all of the shock-surface of the star can efficiently compress CRs, we assumed a cone with an total opening angle of $\theta=$ \ang{90} for the size of the compressing region. As a result, all emission was scaled by $\frac{1-\cos(\theta/2)}{2}\approx0.15$ compared to a fully spherical emission geometry.

The model predicts the observed radio emission reasonably well for BSBD43 (see figure \ref{fig:BD43_compression}) and is at same time not violating the observational upper-limits on the gamma-ray emission. It has to be noted, nevertheless, that the hadronic gamma-ray emission represents the most optimistic scenario.
Any lower ambient-medium density would reduce the hadronic gamma-ray emission while leaving the radio and inverse-Compton emission unchanged, as long as the shock compression-ratio stays the same. 

No reasonable set of parameters is able to reproduce the radio emission from BSBD60 in this scenario.
This is mainly on account of the smaller separation between the star and the bow shock in this case.
The small size of the emission region requires unreasonably large magnetic fields ($B_0\geq \SI{150}{\mu G}$) and high compression factors ($r\geq40$). 

%%%%%%%%%%%%%%%%%%%%%%%%%%%%%%%%%%%%%%%
% 06 - CONCLUSIONS
%%%%%%%%%%%%%%%%%%%%%%%%%%%%%%%%%%%%%%%
\section{Conclusions}\label{sec:concl}
We report two new, high-resolution radio observations of bow shocks from runaway stars, \bdforty (BSBD43) and \bdsixty (BSBD60).
We observed extended emission from both sources with the VLA and Effelsberg telescopes, at \SIrange{4}{12}{GHz} and \SIrange{4}{8}{GHz} respectively.
For BSBD43 we add higher frequency data to the results of \citeadsalias{2010A&A...517L..10B}, \citeadsalias{2021MNRAS.503.2514B}, while for BSBD60 we present the first radio data since the 1970s.
We then combined interferometric and single-dish data with the \textit{feather} technique, to further probe the nature of the emission of the bow shocks.

We have detected, for the first time, non-thermal radio emission from BSBD60, confirming the suggestion of \citetads{2019A&A...625A...4G} that this bow shock could be a bright non-thermal emitter.
The emission traces a portion of the edge of the bubble very well and the brightest region is the photoionised pillar projected behind the bubble.
This pillar has spectral index closest to TE whereas the projected rim of the bubble has more negative spectral index.
Besides its larger distance, BSBD60 appears quite compact compared to BSBD43, and so the inclusion of the Effelsberg data in the feathered dataset does not change the results significantly.
Further interferometric radio observations at lower frequencies are needed to better characterise the emission and cleanly separate thermal and non-thermal emission.

For BSBD43 our results are less conclusive: including only the VLA data we find spectral indices that are so negative that they cannot be physical and must indicate some systematic loss of flux at high frequencies.
Comparing the Effelsberg and VLA flux levels we see that this is indeed the case, and the feathered dataset shows a mildly negative spectral index that could be consistent with TE or a combination of strong TE and weak NTE.
The overall morphology of the VLA data is consistent with previous observations at lower frequency by \citeadsalias{2010A&A...517L..10B} and \citeadsalias{2021MNRAS.503.2514B}.
This suggests it is possible that the VLA observations are picking out small non-thermal emitting regions, which are concealed once the large-scale TE from Effelsberg is added, but this cannot be confirmed since we are missing flux at higher frequencies.
Lower spatial-resolution interferometric data are needed to test this hypothesis.

Finally, we modelled the source of the NTE in the two bow shocks with simple one-zone models of (i) high-energy electrons from shock acceleration at the wind termination shock, and (ii) emission from Galactic cosmic rays compressed by the radiative forward shock of the bow shock.
For the termination-shock model we find that it can explain the radio emission and is consistent with upper limits in X-rays and $\gamma$-rays.
Using arguments based on the total energy available in non-thermal electrons, we show that the non-thermal X-ray and $\gamma$-ray emission could be well below current observational upper limits, for both sources.
For the shock-compression model we find that it could explain the radio flux levels for BSBD43, and predicts IC $\gamma$-ray emission about 0.5 dex below the FERMI upper limits.
In the case of BSBD60 the bow shock is too small and the non-thermal radio emission cannot be explained by this model.

%%%%%%%%%%%%%%%%%%%%%%%%%%%%%%%%%%%%%%%
% ACKNOWLEDGEMENTS
%%%%%%%%%%%%%%%%%%%%%%%%%%%%%%%%%%%%%%%
\begin{acknowledgements}
MM acknowledges funding from the Royal Society Research Fellows Enhancement Award 2017 (17/RS-EA/3468).
JM acknowledges funding from a Royal Society-Science Foundation Ireland University Research Fellowship (14/RS-URF/3219, 20/RS-URF-R/3712).
DZ and RB acknowledge funding from an Irish Research Council Starting Laureate Award (IRCLA\textbackslash2017\textbackslash83).
This work is partly based on observations with the 100-m telescope of the MPIfR (Max-Planck-Institut für Radioastronomie) at Effelsberg. JAT acknowledges funding from the UNAM DGAPA PAIIT (Mexico) project IA101622 and the Marcos Moshinsky Foundation (Mexico).
The National Radio Astronomy Observatory is a facility of the National Science Foundation operated under cooperative agreement by Associated Universities, Inc.
This publication makes use of data products from the Wide-field Infrared Survey Explorer, which is a joint project of the University of California, Los Angeles, and the Jet Propulsion Laboratory/California Institute of Technology, funded by the National Aeronautics and Space Administration.
This work has made use of data from the European Space Agency (ESA) mission
{\it Gaia} (\url{https://www.cosmos.esa.int/gaia}), processed by the {\it Gaia}
Data Processing and Analysis Consortium (DPAC,
\url{https://www.cosmos.esa.int/web/gaia/dpac/consortium}). Funding for the DPAC
has been provided by national institutions, in particular the institutions
participating in the {\it Gaia} Multilateral Agreement.
The authors would like to thank the referee for their careful reading of our paper and constructive comments that improved the manuscript.
Vasilii V.~Gvaramadze very sadly passed away during the writing this paper; we will miss him.
\end{acknowledgements}
%%%%%%%%%%%%%%%%%%%%%%%%%%%%%%%%%%%%%%%
%%%%%%%%%%%%%%%%%%%%%%%%%%%%%%%%%%%%%%%

\bibliographystyle{aa} % style aa.bst
\bibliography{bib-new} 

\appendix
\section{Analysis including the H~\textsc{II} regions surrounding the sources}
\begin{figure}[!ht]
    \centering
    \includegraphics[width=\linewidth]{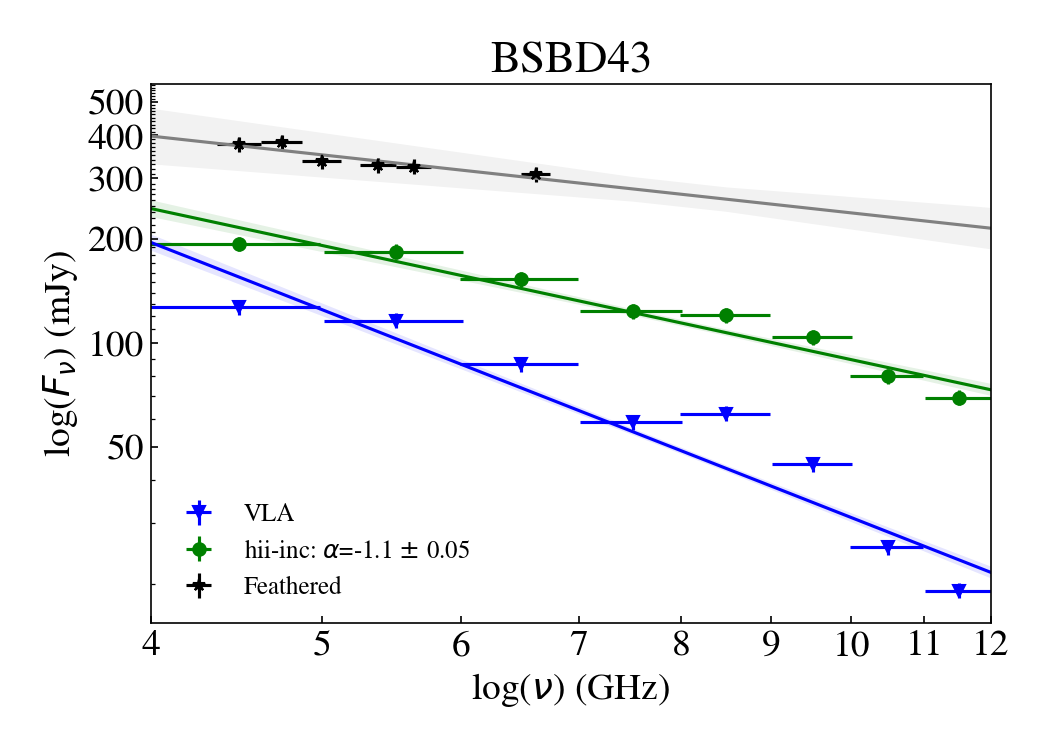}\\
    \includegraphics[width=\linewidth]{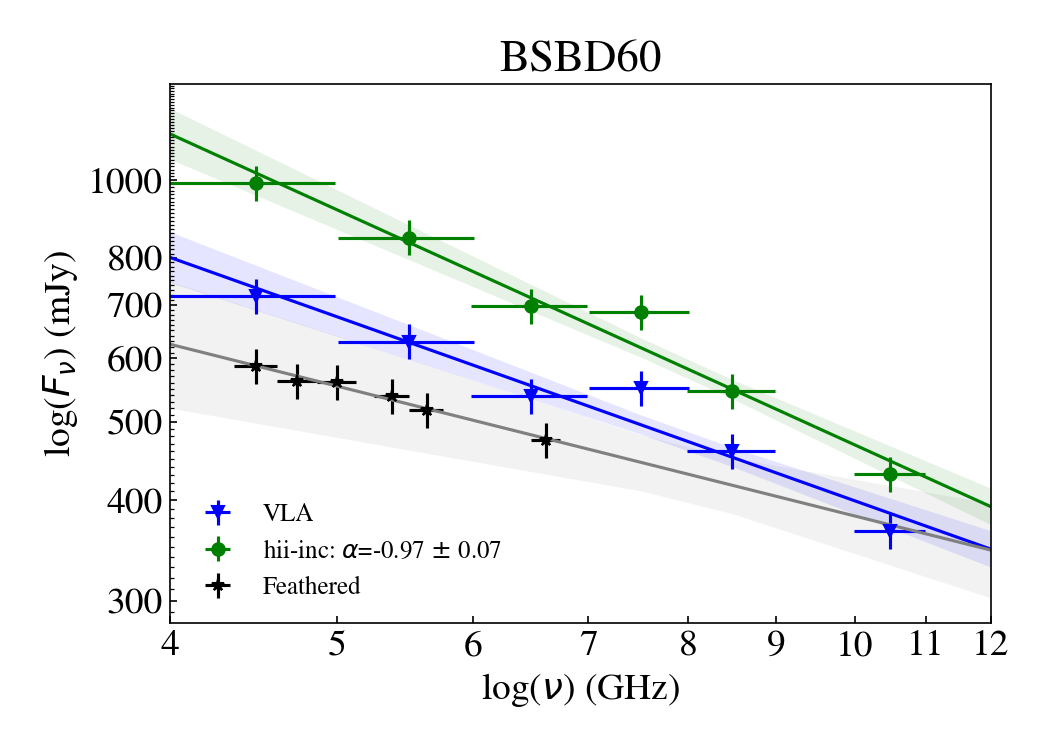}
    \caption{SED with H~\textsc{ii} area above source included in VLA and feathered data.}
    \label{fig:app_01}
\end{figure}

It is worth discussing the difference it would make to our results if the H~\textsc{ii} region above each of our sources was accounted for in our calculations for the spectral index.
As it is mentioned in Sec. \ref{subsubsec:ifrd}, we assumed for our calculations that only the bow shock is contributing to the total emission.
It is, however, possible that the H~\textsc{ii} region located at the northern part of our field of view can have an additional effect to the final results, especially given the large beam size of the Effelsberg data. 
Therefore, we included these sources (again for regions with flux above 5$\sigma$ in the VLA map) and made an additional calculation of the spectral index for our data.
The results are presented in Fig. \ref{fig:app_01}.
For BSBD43, the spectral index becomes less negative, as expected because we add a presumably thermal source to the total.
For BSBD60, the spectral index became more negative, likely because the VLA spectral index for these bright-rimmed clouds is strongly negative, albeit very uncertain (see Fig.~\ref{fig:bd_spindx_all}).
Here the results are inconclusive.

\section{uv-range}
\begin{figure*}
    \centering
    \includegraphics[width=0.5\linewidth,trim={2 0 0 2},clip]{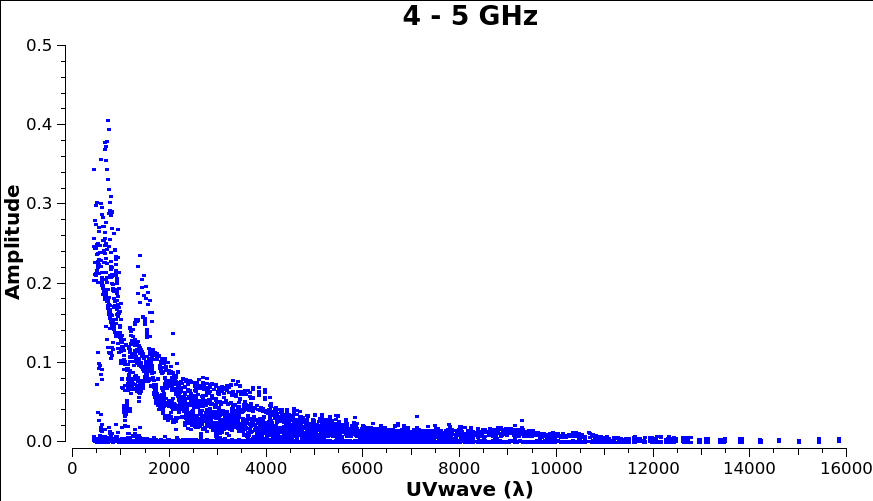}\hfill
    \includegraphics[width=0.5\linewidth,trim={2 0 0 2},clip]{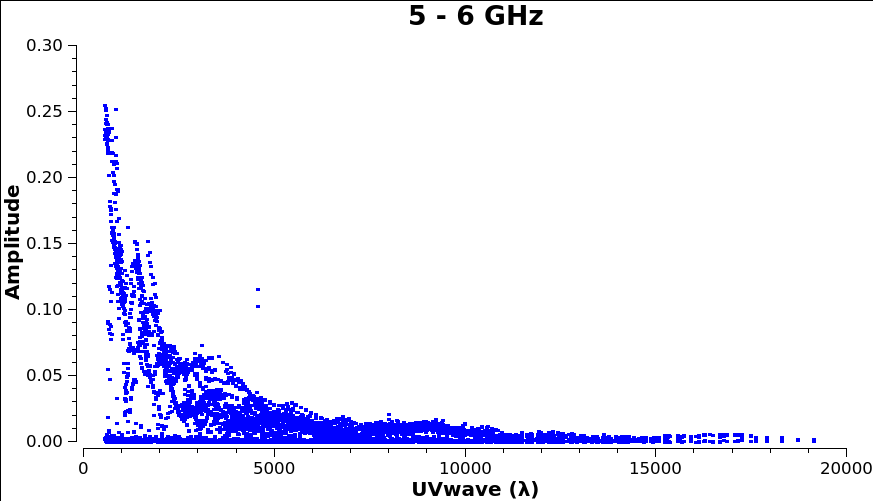}\\
    \vspace{0.2cm}
    \includegraphics[width=0.5\linewidth,trim={2 0 0 2},clip]{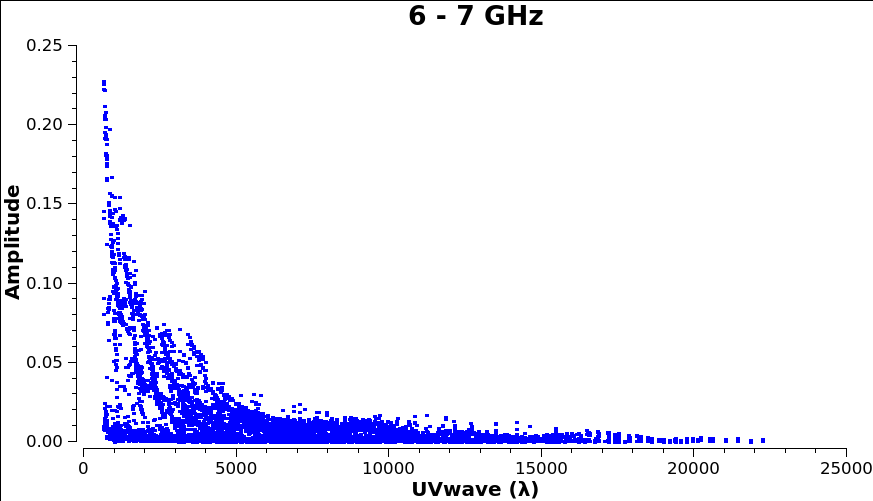}\hfill
    \includegraphics[width=0.5\linewidth,trim={2 0 0 2},clip]{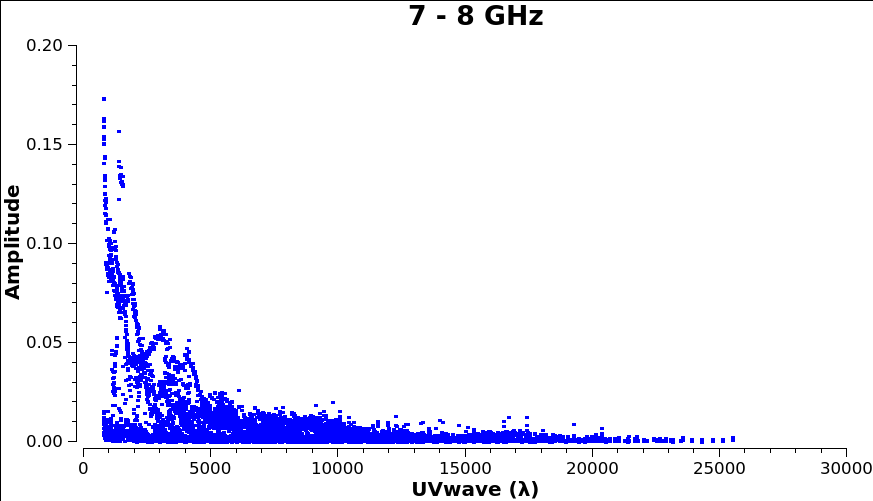}\\
    \vspace{0.2cm}
    \includegraphics[width=0.5\linewidth,trim={2 0 0 2},clip]{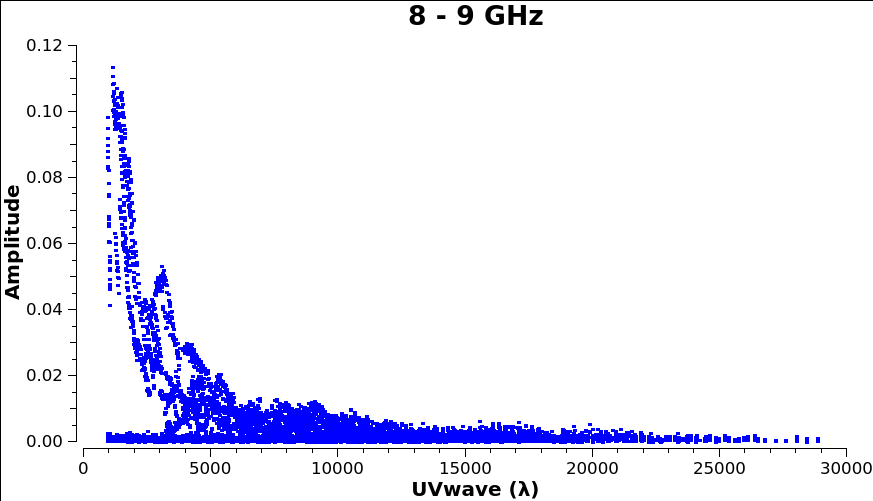}\hfill
    \includegraphics[width=0.5\linewidth,trim={2 0 0 2},clip]{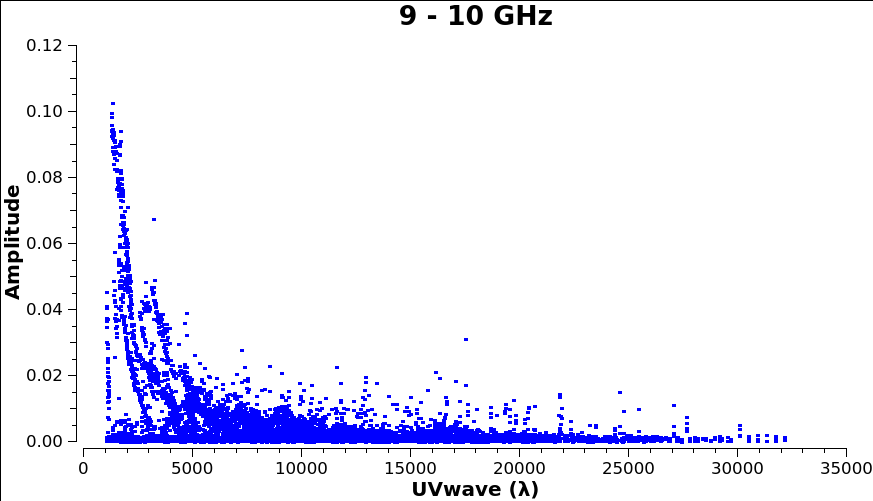}\\
    \vspace{0.2cm}
    \includegraphics[width=0.5\linewidth,trim={2 0 0 2},clip]{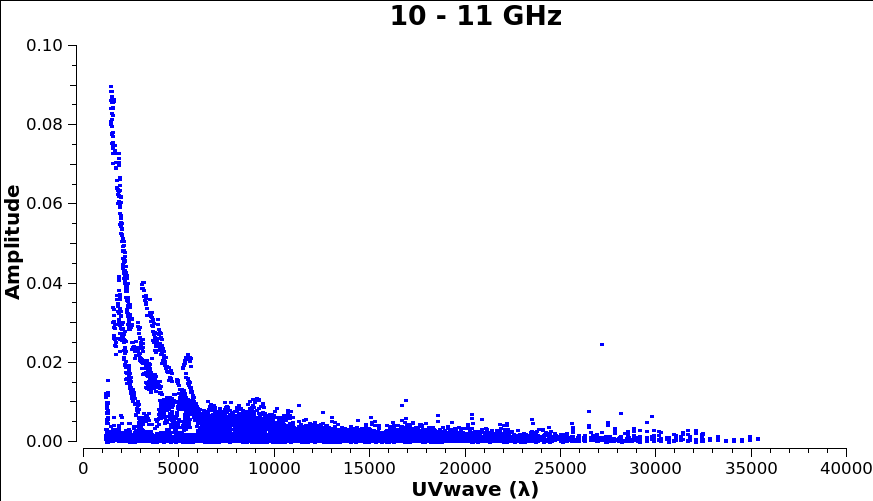}\hfill
    \includegraphics[width=0.5\linewidth,trim={2 0 0 2},clip]{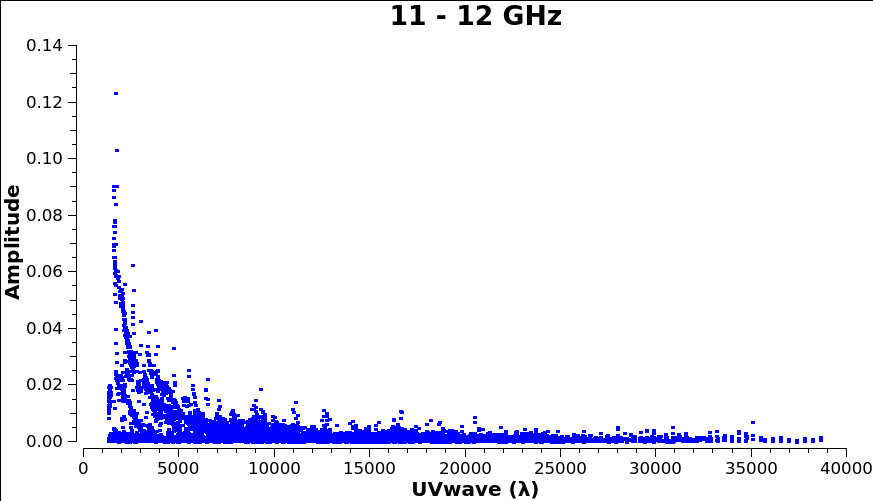}
    \caption{Intensity amplitude vs. $uv-$wave (measured in the observed wavelength). Each plot represents a \SI{1}{GHz}-wide band of our BSBD43 data.}
    \label{fig:app_02}
\end{figure*}

The $uv$-range for each \SI{1}{GHz}-wide band of BSBD43 is show in Fig. \ref{fig:app_02}.
The equivalent plots for BSBD60 are omitted since they were similar, only different in amplitude.
As it can be seen, all of them have a significant amount of information, thereby justifying the investigation of flux in the whole bandwidth.

\end{document}